\documentclass[conference]{IEEEtran}
\IEEEoverridecommandlockouts
\usepackage{amsmath,amssymb,amsfonts}
\usepackage{algorithmic}
\usepackage{graphicx}
\usepackage{textcomp}
\usepackage[table]{xcolor}
\def\BibTeX{{\rm B\kern-.05em{\sc i\kern-.025em b}\kern-.08em
    T\kern-.1667em\lower.7ex\hbox{E}\kern-.125emX}}

\usepackage{soul}
\sethlcolor{yellow!50}
\usepackage{subcaption}

\usepackage{comment}
\usepackage{multirow}
\usepackage[backend=biber]{biblatex}
\begin{document}

\title{PM2Lat: Highly Accurate and Generalized Prediction of DNN Execution Latency on GPUs
%\thanks{Identify applicable funding agency here. If none, delete this.}
}

\author{\IEEEauthorblockN{1\textsuperscript{st} Truong-Thanh Le}
\IEEEauthorblockA{\textit{Department of Informatics} \\
\textit{University of Oslo}\\
Oslo, Norway \\
truongl@ifi.uio.no}
\and
\IEEEauthorblockN{2\textsuperscript{nd} Hoang-Loc La}
\IEEEauthorblockA{\textit{Department of Computer Science} \\
\textit{The Arctic University of Norway}\\
Tromsø, Norway \\
hoang.l.la@uit.no}
\and
\IEEEauthorblockN{3\textsuperscript{rd} Amir Taherkordi}
\IEEEauthorblockA{\textit{Department of Informatics} \\
\textit{University of Oslo}\\
Oslo, Norway \\
amirhost@ifi.uio.no}
\and
\IEEEauthorblockN{4\textsuperscript{th} Frank Eliassen}
\IEEEauthorblockA{\textit{Department of Informatics} \\
\textit{University of Oslo}\\
Oslo, Norway \\
frank@ifi.uio.no}
\and
\IEEEauthorblockN{5\textsuperscript{th} Phuong Hoai Ha}
\IEEEauthorblockA{\textit{Department of Computer Science} \\
\textit{The Arctic University of Norway}\\
Tromsø, Norway \\
phuong.hoai.ha@uit.no}
\and
\IEEEauthorblockN{6\textsuperscript{th} Peiyuan	Guan}
\IEEEauthorblockA{\textit{Department of Informatics} \\
\textit{University of Oslo}\\
Oslo, Norway \\
peiyuang@ifi.uio.no}
}

\maketitle

\begin{abstract}
We present PM2Lat, a fast and generalized framework for accurately predicting the latency of deep neural network models on GPUs, with special focus on NVIDIA. Unlike prior methods that rely on deep learning models or handcrafted heuristics, PM2Lat leverages the Single-Instruction-Multiple-Thread architecture of GPUs to model execution time of DNN models. %By analyzing kernel-level throughput and separating compute- and memory-intensive operations, PM2Lat demonstrates that simple interpolation is sufficient to achieve high prediction accuracy.
%By diving deeper into the operation modeling of GPUs including computational behavior and memory access pattern, 
%we focus on the differences in performance among types of GPU kernel even if they serve the same purposes. Hence, PM2Lat provides ability of achieving stably low prediction error rate regarding of data types and hardware.
First, we dive into fine-grained GPU operation modeling by studying computational behavior and memory access patterns. After identifying these characteristics, we found that different GPU kernels exhibit significant performance disparities, even when serving the same purpose. Hence, the core idea of PM2Lat is to differentiate kernels based on their configurations and analyze them accordingly. This kernel-aware modeling enables PM2Lat to achieve consistently low prediction error across diverse data types and hardware platforms.
In addition, PM2Lat generalizes beyond standard matrix multiplication to support complex GPU kernels such as Triton, Flash Attention, and Cutlass Attention. Experimental results show that PM2Lat consistently achieves error rates below 10\% across different data types and hardware platforms on Transformer models, outperforming the state-of-the-art NeuSight by 10–20\% for FP32 and by at least 50\% for BF16. When applying to diverse kernels, the error rate is maintained at 3-8\%.
%Experimental results show that PM2Lat stabily achieve error rate of less than 10\% across difference datatypes and hardware on Transformer models, outperform the state-of-the-art NeuSgith by 10-20\% with FP32, and by at least 50\% with BF16. 
%outperforms the state-of-the-art NeuSight by 10–20\% in accuracy on Transformer models with FP32 data type, while maintaining error rates of 3–8\% across diverse kernels. With BF16 datatype, 
\end{abstract}

\begin{IEEEkeywords}
Matrix Multiplication, Attention, Transformers, Computation Intensive, Memory Intensive, DNN Latency, CUDA
\end{IEEEkeywords}

\section{Introduction}\label{sec.Intro}

Accurate prediction of deep neural network (DNN) execution latency on GPUs has become a critical requirement for modern AI systems. Applications such as distributed training/inference, and large-scale deployment rely heavily on these predictions to guide resource allocation, workload partitioning, and performance optimization. As models grow to billions of parameters and operate across heterogeneous environments, even small latency misestimations can lead to inefficient scheduling and degraded user experience.
For example, systems like ChatGPT and Copilot rely on cloud-based GPU clusters and load balancing to deliver low-latency responses at scale \cite{yan2024exploring,zhang2024edgeshard}.
However, in such applications, even small delays matter---milliseconds saved per request can significantly improve overall performance when millions of queries are processed \cite{le2024optimal}.
%To enable such optimizations, developers need reliable tools to predict execution time across heterogeneous environments spanning Edge, Fog, and Cloud \cite{zhang2024edgeshard}. These predictions guide workload partitioning and resource allocation, which is challenging given the diverse hardware capabilities of different devices.

The need for accurate latency prediction becomes even more critical during the design and optimizing phase of LLMs/DNNs (fine-tune, distillation, pruning, etc), where developers may use Neural Architecture Search (NAS)~\cite{cai2018proxylessnas}. 
%, which explores architectures balancing accuracy and efficiency on resource-constrained devices to meet Quality-of-Service requirements such as real-time operation and low latency service \cite{pham2018efficient, tan2019mnasnet}.
%In the context of LLMs, NAS is used to fine-tune architectural components such as attention mechanisms, feedforward layers, and embedding dimensions, each of which can significantly impact both model quality and runtime performance.
%The challenge becomes even more pressing in Neural Architecture Search (NAS), where developers must evaluate millions of model designs to find the most efficient one \cite{pham2018efficient, tan2019mnasnet}. 
%However, each candidate architecture may vary in layer types, input/output dimensions, batch sizes, and sequence lengths, leading to an exponential growth in the number of configurations that must be assessed. Hence, evaluating these configurations is \textbf{computationally expensive due to the sheer volume of possible designs}, often resulting in \textbf{significant time overhead}. 
While it is often argued that in NAS, regarding the latency factor, we only need sufficiently accurate estimation to guide DNN topology selection, this assumption holds mainly for applications with low time sensitivity.
In NAS for latency-critical scenarios---such as real-time services or resource-constrained deployments involving model partitioning, data parallelism, or collaborative execution across Edge–Fog–Cloud infrastructures---%even small prediction errors can lead to suboptimal architectures, reducing significantly  Quality-of-Service, as mentioned above??. 
higher latency prediction error can lead to the suboptimal model architectures as the objectives mainly relies on the trade-off between latency and accuracy of the model \cite{cai2018proxylessnas}.
Therefore, highly precise latency prediction is essential to ensure both performance and reliability in practical systems.
%Typically, NAS fine-tunes components like attention mechanisms and feedforward layers.
%However, each candidate architecture can vary in dimensions, batch size, sequence length, and accuracy characteristics, resulting in an exponential number of possible configurations.
%, but each candidate varies in dimensions, batch sizes, sequence lengths and of course the accuracy impact, creating an exponential number of configurations. 
However, evaluating a sheer number of candidate architectures when running NAS is \textit{computationally expensive} and \textit{incurs significant time overhead} \cite{cai2018proxylessnas}.

A common approach to reduce the time overhead of predictions is to precompute latency for all possible settings and store them in a cache for future re-use \cite{akhauri2024latency}. This allows for fast lookups during NAS or deployment, significantly cutting down runtime overhead. However, this step is far from trivial. The sheer number of configurations, combined with the need for accurate predictions, makes it extremely time-consuming. For instance, even if each prediction requires only 1 $ms$, the total time can accumulate quickly to hours, days or even months.

%Model designs can vary in layer types, input/output sizes, and batch settings. Even if each latency prediction requires only one millisecond, the total time can accumulate quickly. 
%For example, in a Transformer model with 14 choices for input/output features, batch sizes ranging from 1 to 256, and sequence lengths from 64 to 8192, the number of configurations for just one matrix multiplication (MatMul) layer exceeds 400 million possibilities.
%On a single target device, evaluating these configurations at just one millisecond per measurement would require nearly five days of continuous processing.
%With multiple devices involved, the computational workload multiplies, making \textit{fast and scalable latency prediction not just helpful, but essential.}

Traditionally, developers rely on proxy metrics for all type of DL layers, such as the number of parameters or FLOPs (floating-point operations), to estimate performance \cite{krishnakumar2022bench}. These metrics are reliable enough when being applied on memory-intensive layers, where the latency mostly relies on memory bandwidth. However, in compute-intensive layers, these metrics often miss important details \cite{la2025kernellevelenergyefficientneuralarchitecture,colin2022adeeperlook}, such as memory access patterns, kernel-level efficiency, and computation-memory bottlenecks \cite{ye2025flashinfer}. A model that appears efficient in theory may still perform poorly in practice, especially if it struggles with memory access patterns.% or uses GPU kernels that are not well optimized. %This gap between theoretical estimates and actual performance can lead to poor performance and wasted resources.
%This gap can lead to poor performance and wasted resources.

Recent research efforts, such as NeuSight \cite{NeuSight} and Habitat \cite{geoffrey2021habitat}, have addressed this problem by proposing appropriate DL models to predict latency.
Although they achieve high accuracy, they induce considerable time overhead.
%they are slow and resource-intensive.
Specifically, due to relying on DNNs with many layers and parameters, these approaches are considered “heavy”.
%, especially when applied to large batches of configurations in NAS.
%even with dedicated GPUs.
%for acceleration.
%They comes with non-negligible time overhead to run, 
%During inference, these models can introduce non-negligible overhead, 
As a result, the prediction process may not be fast enough when using NAS preprocessing for model design for resource constrained devices, where millions or even billions of latency predictions must be made quickly \cite{ghebriout2024harmonic}. 
%latency-sensitive application design, e.g.,
%using NAS for resource constrained devices . 
%or real-time deployments, where millions or even billions of latency predictions must be made quickly. 
%In addition to runtime cost, these methods also add complexity in terms of model training, tuning, and integration into existing systems. 
Moreover, state-of-the-art methods primarily focus on the 32-bit floating-point (FP32) data type, while modern LLMs and DNNs are increasingly using BFloat16 (BF16) or even smaller data types. However, FP32 operations run on CUDA cores, whereas BF16 computations are executed on Tensor Cores, which exhibit significant performance differences. This raises concerns about the generalizability of existing approaches when being applied to different data types.
Furthermore, many of these methods target specific GPU kernel types, which limits their generalizability to newer or more complex workloads.
%related work on FLOAT32/16??

%This brings us to the key question: \textbf{can we predict execution latency of a DNN in a highly accurate manner without relying on heavy models with long processing times?}
This leads to a key question: \textit{Can we accurately predict the execution latency of a DNN without relying on heavyweight models that incur substantial processing time?}

In this work, we introduce PM2Lat, a novel method for predicting the latency of DL models that is fast and general enough to work across many types of GPU kernels, data types and devices. 
PM2Lat builds on a key observation:\textit{ GPU performance is not consistent across different kernels even when kernels serve the same purpose, because implementation details vary}. For example, recent NVIDIA GPUs provide over 25 different kernels for FP32 MatMul. Although these kernels perform the same number of FLOPs, their memory access patterns differ significantly, making latency prediction based solely on FLOPs unreliable.
In addition, \textit{under the Single-Instruction-Multiple-Thread (SIMT) execution model of GPUs, memory access patterns and other hidden characteristics exhibit consistent behavior}. In other words, \textit{SIMT ensures that threads execute instructions in lockstep with minimal divergence without caring about hidden characteristics}. 
Hence, instead of attempting to model memory access patterns using limited kernel configuration information—which is challenging due to the lack of APIs and closed-source constraints—we found it more effective to differentiate kernels. By analyzing each kernel accordingly, we achieve better results while avoiding unseen factors.
%Hence, we found that instead of attempting to model memory access patterns using limited kernel configuration information, which is challenging due to the lack of APIs and closed-source constraints, we achieve better results by differentiating kernels and analyzing them accordingly, avoiding unseen factors. 
This approach enables PM2Lat to generalize across data types, layer types, and hardware platforms.
%The generalization stems from a key insight: \textit{under the Single-Instruction-Multiple-Thread (SIMT) execution model of GPUs, memory access patterns and other hidden characteristics exhibit consistent behavior}. In other words, \textit{SIMT ensures that threads execute instructions in lockstep with minimal divergence without caring about hidden characteristics}. Leveraging this property allows PM2Lat to avoid complex models for latency prediction to capture hidden features.

Our contributions are summarized as follows:
\begin{itemize}
    \item We develop a \textit{novel, lightweight, highly accurate}, and \textit{generalized} method using kernel differentiation and SIMT characteristics for predicting the latency of both \textit{Computation-intensive} and \textit{Memory-intensive} kernels on NVIDIA GPUs. Building on this, we develop a unified approach to estimate the latency of entire DNN models, as described in Section~\ref{sec.ProposedMethod}.
    
    \item We conduct extensive experiments across multiple NVIDIA GPUs and demonstrate that our proposed method, \textit{PM2Lat, stably achieves error rate of lower than 10\%}. In addition, we found that PM2Lat significantly outperform NeuSight on BF16 datatype, with at least 50\% difference (cf. Section~\ref{sec.Evaluation}).

    \item We also apply our method to two representative applications to demonstrate PM2Lat’s applicability: \textit{i)} Model partitioning for distributed inference and collaboration across heterogeneous Edge devices; \textit{ii)} Prediction latency for NAS preprocessing step (cf. section \ref{subsec.Eval.Applications}). Through these use cases, we highlight the importance of a \textit{fast and accurate} prediction method.
    
\end{itemize}

\section{Related Work} \label{sec.RelatedWork}

Latency prediction for DNNs has become a critical component in optimizing model deployment, especially in large-scale and distributed systems. Two of the most prominent methods in this space are NeuSight \cite{NeuSight} and Habitat \cite{geoffrey2021habitat}, both of which aim to estimate the latency of individual GPU kernels and then aggregate these estimates to predict the total model latency.% Comments on other related works can be found in Appendix \ref{sec.APD.relatedwork}.

Habitat is a runtime-based framework designed to estimate training time across different GPU architectures. It combines wave (see section \ref{sec.ProposedMethod}) scaling techniques with a pre-trained multi-layer perceptron (MLP) to achieve high prediction accuracy. One of its strengths lies in its ability to handle kernel-varying layers—those with multiple implementations—using only a single iteration of training data. This reduces the need for extensive benchmarking, making it more practical for early-stage evaluations. For layers with fixed implementations, Habitat executes them on a reference GPU and scales the results based on the target device’s specifications. This approach helps developers make informed decisions about model architecture and deployment strategies.

NeuSight, on the other hand, offers a more detailed analysis of MatMul layers, including fully connected and batched MatMul operations. It uses a tile-based method to divide output matrices into smaller blocks, each with distinct performance characteristics. To predict latency, NeuSight trains deep neural networks to estimate GPU utilization based on the number of execution waves—a key factor affecting runtime. By matching new input shapes to a large dataset of known tile configurations, NeuSight can select the most suitable tile size and predict latency without actual execution. This method has shown strong accuracy, especially for complex models like GPT-3 on high-end GPUs.

Despite their strengths, both methods rely heavily on DL models (MLPs or DNNs) for prediction. This introduces significant computational overhead, particularly during the preprocessing phase of NAS, where millions of configurations must be evaluated. The time and resources required to train and run these models can slow down architecture search and deployment decisions. In addition, both methods only evaluate their efficiency using FP32 datatypes, assuming uniform characteristics across different precision levels. However, this assumption raises concerns about its correctness across different precision formats.

In contrast, our proposed method, PM2Lat, takes a different approach. 
%Instead of relying on DL models, PM2Lat uses lightweight regression and interpolation techniques to predict latency. 
%It focuses on modeling the relationship between FLOPs and execution time, while also accounting for throughput variations and kernel-specific behaviors. 
Instead of relying on DL models, we simplify latency prediction through a mathematical estimation framework based on kernel-level differentiation and the characteristics of SIMT architectures.
This approach allows PM2Lat to adapt seamlessly across different precision formats and device architectures, while delivering significant improvements in prediction speed, making it a robust solution for latency prediction in diverse deployment scenarios.
%By avoiding the complexity of DNN-based predictors, PM2Lat achieves faster prediction times and better scalability, making it more suitable for real-world deployment and large-scale NAS.

\section{Proposed method} \label{sec.ProposedMethod}
%We introduce PM2Lat, a lightweight and accurate framework for predicting the execution latency of DNNs on GPUs, with specifically focus on NVIDIA. 
The core idea behind PM2Lat is to model the relationship between FLOP count and execution time for computation-intensive kernels, leveraging the inherent parallelism of GPU architectures. While this concept has been explored in prior work, our observations and modeling approach differentiate PM2Lat from existing methods. We begin with a detailed analysis of MatMul kernels---one of the most widely used and compute-intensive operations in modern DL models---and then generalize to other kernel types.
Specifically, we treat each unique combination of MatMul configurations (e.g., algorithms, tiling strategies, and precision modes) as a distinct kernel. We then empirically verify that the execution time of these kernels exhibits a strong rational correlation with their FLOP counts, consistent with the SIMT architecture. %Hence, focusing solely on MatMul as a representative of computation-intensive kernels does not restrict the applicability of our approach. Instead, it provides a solid foundation for generalizing latency prediction to other GPU kernels that are also computation-intensive.

For memory-bound utility layers such as Pooling, ReLU, GeLU, Add, Mul, and Dropout, PM2Lat employs proxy metrics like operation counts and memory access volume---as they are still reliable due to memory bandwidth dependent--- and then applying linear regression to estimate latency. These metrics are collected using profiling tools and scaled appropriately across devices, \textit{rather than relying on theoretical models}, which often diverge from real-world performance due to variations in code design and implementation.

To estimate the total latency of a DNN model, PM2Lat aggregates the predicted latencies of all layers, assuming sequential execution of CUDA kernels. This approach aligns with the execution behavior observed in frameworks like NeuSight \cite{NeuSight} and Habitat \cite{geoffrey2021habitat}.

Before diving into the details of our method, we first review the GPU architecture, kernel execution behavior and NeuSight's limitations that form the foundation of our latency prediction model. 
%We also discuss the limitations of NeuSight, particularly its lack of generalizability and high computational overhead, to motivate our design choices in PM2Lat. 
%Some technical terms and optimization techniques referenced in this section are explained in Appendix \ref{sec.APD.MatMulOptim} for clarity.

\subsection{Theory background} \label{subsec.Theory}

\begin{figure}
    \centering
    \includegraphics[width=0.7\linewidth]{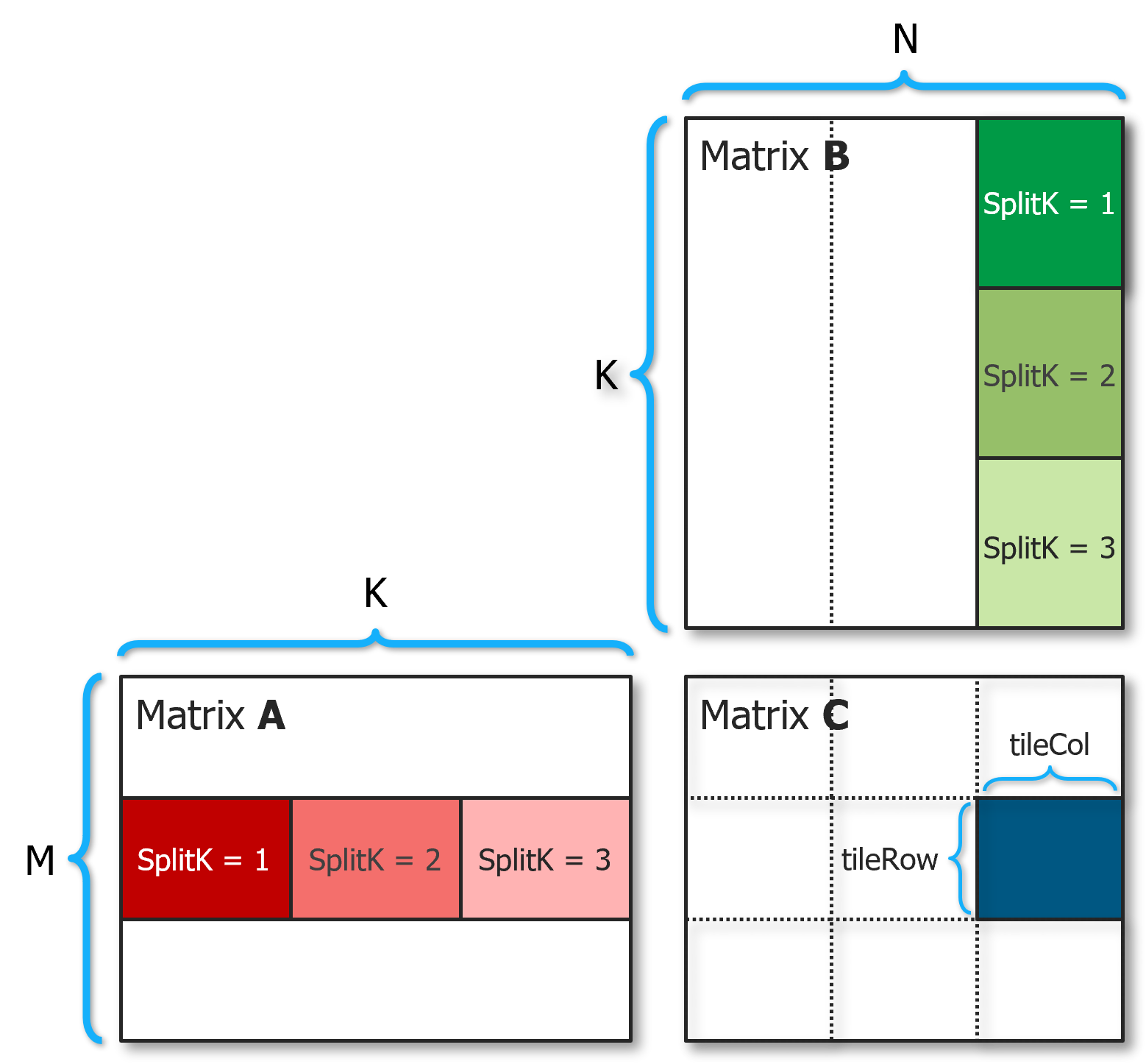}
    \caption{Example of Tile-Based GEMM.}
    \label{fig:TileGEMM}
\end{figure}

\textbf{Understanding GPU Behavior for Latency Prediction:} NVIDIA GPUs are designed for high-throughput parallelism using a SIMT architecture. \textit{This design allows thousands of lightweight threads to execute the same instruction across different data elements simultaneously}. As a result, operations like MatMul and convolution can be executed efficiently, and their latency often scales linearly with the number of operations—especially for compute-intensive kernels.

However, previous methods that rely solely on this linear relationship have shown limited accuracy \cite{qi2017paleo,wang2024latte}. This is because other factors—such as memory access patterns, kernel configuration, and hardware constraints—can significantly affect prediction accuracy. To address these limitations, PM2Lat performs a deeper analysis of kernel behavior, starting with MatMul and Utilities layers, and then generalize to other GPU kernels like Triton, Flash Attention and Cutlass Attention. It is important to note that most training and inference workloads today rely heavily on NVIDIA GPUs. Their widespread use in DL systems is driven by strong hardware capabilities and a mature software ecosystem that supports efficient parallel computation \cite{deepseekai2025deepseekr1incentivizingreasoningcapability}. This includes advanced methods developed for LLMs, such as Flash Attention \cite{dao2023flashattention2} and Cutlass Attention \cite{xFormers2022}, which are specifically optimized for NVIDIA architectures. For this reason, our work focuses exclusively on NVIDIA GPUs.
Although AMD GPUs share a similar SIMT architecture,
%and may exhibit comparable behavior
evaluating their performance is left for future work.

\textbf{MatMul and GPU Optimization:} MatMul is a fundamental operation in DL, used extensively in dense layers, convolution and attention mechanisms. It computes the dot product between rows and columns of input matrices, producing an output matrix that captures relationships between features. We also acknowledge that many recent DNN approaches employ sparse MatMul to reduce computation. However, sparse computation exhibits significantly different performance characteristics compared to dense MatMul, including irregular memory access patterns, load imbalance, and library‑specific optimization challenges. Therefore, analyzing sparse MatMul behavior is beyond the scope of this work, and we leave it for future investigation.

On a typical GPU, MatMul is optimized using a tile-based strategy. Large matrices are divided into smaller tiles that fit into shared memory, as shown in Figure \ref{fig:TileGEMM}. Each tile is processed by a block of threads, which improves data reuse and reduces global memory access. This strategy enhances scalability and throughput, especially for large models.

The execution of MatMul on GPUs involves multiple waves---the sequential batches of thread blocks. Due to limited shared memory and a finite number of streaming multiprocessors (SMs), only a subset of tiles can be processed at a time, called a wave. Once a wave completes, the next is scheduled. The total number of waves depends on matrix size, tile configuration, and available hardware resources.

NVIDIA provides two major libraries for MatMul: cuBLAS and CUTLASS. cuBLAS is a highly optimized library available across all NVIDIA GPUs, while CUTLASS offers customizable CUDA C++ templates for high-performance MatMul operations on newer devices. These libraries support various optimization techniques, including splitK, reduction scheme, swizzling \cite{nvidia2025cutlass}.
%\begin{itemize}
%    \item \textbf{splitK}: Instead of computing a full tile, this technique divides a tile into K smaller tiles along with K-dimension (Figure \ref{fig:TileGEMM}) for better paralellism. 
%    \item \textbf{Reduction schemes}: when applying splitK, there must be a method to aggregate the results of each splitK tile.
%    \item \textbf{Swizzling}: Improved memory access patterns and load balancing.
%\end{itemize}
These techniques are integrated into the unified cuBLAS API, which automatically selects the best kernel configuration based on hardware and workload characteristics. Detailed explanations of these optimizations can be found in the official documentation provided by NVIDIA \cite{nvidia2025cutlass}. We will not analyze the impact of each method individually, as we assume that each combination of these methods operates as a single kernel, executing in a consistent manner throughout its lifetime, therefore eliminating the hidden factors. It should be noted that cuBLAS can internally invoke CUTLASS kernels for better performance.

\textbf{Utility Layers:} Utility layers in DNNs—such as activation and normalization functions—do not contain learnable parameters but play a key role in shaping the data flow. These layers typically involve lightweight, element-wise or reduction-based computations that are highly parallelizable. However, their latency is influenced more by memory bandwidth than by computational complexity. Access patterns to DRAM, L2 cache, and L1 cache (within SMs) determine how quickly these operations can be executed \cite{nvidia2025cutlass}.

\subsection{Limitations of SotA Methods} \label{subsec.Limitation}
To motivate the design of PM2Lat, we begin by examining the limitations of NeuSight---the most recent leading framework for DNN latency prediction. While NeuSight demonstrates high accuracy and a deep understanding of GPU execution patterns, several aspects of its methodology limit its efficiency and generalizability.

\textbf{GPU Modeling Gaps:}
NeuSight primarily focuses on theoretical peak FLOPs, DRAM bandwidth, L2 cache size, the number of SMs, and the number of physical cores per SM. However, it overlooks several critical aspects of the GPU architecture, such as L2 bandwidth, L1 bandwidth, and the access patterns between caches and DRAM. 

\begin{figure} 
    \centering \includegraphics[width=0.8\linewidth]{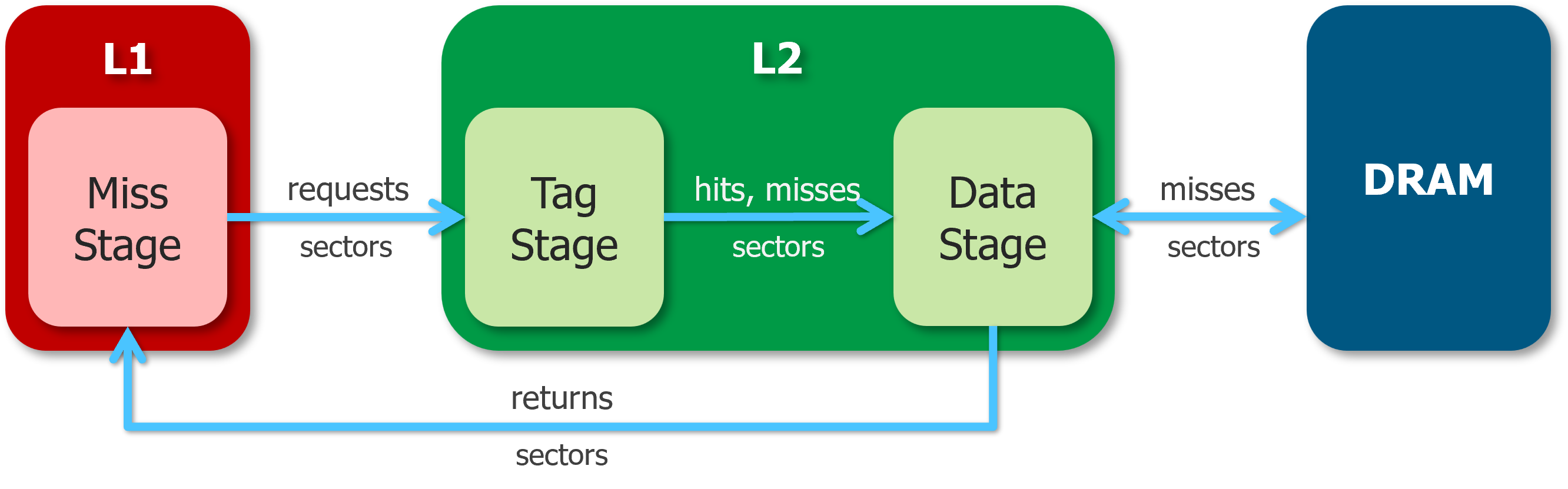} 
    \caption{Memory access flow in NVIDIA GPU} 
    \label{fig:MemMovement} 
\end{figure}

Contrary to the assumption that overall memory bandwidth is solely dependent on DRAM, it is actually a composite of DRAM and high-speed L1 and L2 caches, as illustrated in Figure \ref{fig:MemMovement}. These caches play a vital role in reducing the frequency of DRAM accesses, thereby enhancing effective memory bandwidth. Unfortunately, bandwidth metrics for L1 and L2 caches are not publicly disclosed by GPU manufacturers, making them unobservable and difficult to model accurately.

This missing information becomes particularly problematic when training NeuSight across a diverse set of devices. The absence of key architectural features in the input data can lead to measurable prediction errors between devices. More critically, when NeuSight is applied to unseen hardware---one of its core motivations---this lack of essential data introduces uncertainty. Without these features, there is no guarantee that the prediction error on new hardware can be bounded, which undermines the reliability and generalizability of the model.

Given these limitations, we choose not to model the GPU for deployment on unseen hardware. Instead, we analyze GPU behavior separately for each device. This approach allows us to maximize achievable accuracy while improving the stability and generalization of our predictions, without relying heavily on incomplete or unavailable architectural specifications. Moreover, since these architectural features remain constant for a given device, we treat them as fixed parameters and exclude them from our analysis. 
%{CP version here}
In practice, for newer or newly added devices, we will rerun the full data‑collection and analysis process on the target hardware to obtain the necessary device‑specific characteristics before applying our estimation.

\textbf{Library and Kernel-Level Modeling Gaps:} 
%NeuSight primarily models latency using number of FLOP needed, wave, and GPU specifications (such as theoretical peak FLOPs, DRAM bandwidth, L2 amount, etc.). However, it overlooks key performance differences between GPU libraries such as cuBLAS and CUTLASS---for example, the actual implementation of the kernel can change number of launched wave and FLOP, including control-logic--- which can produce kernels with identical tile sizes but significantly different execution efficiencies due to internal optimizations.
%{CP version here}
NeuSight primarily models latency using theoretical factors such as FLOP counts, wave occupancy, and GPU specifications (e.g., peak FLOPs, DRAM bandwidth, L2 capacity). However, it overlooks critical performance differences introduced by the underlying GPU libraries (e.g., cuBLAS vs. CUTLASS). In practice, the actual kernel implementation can substantially alter the number of launched waves, the effective FLOPs executed, the amount of control‑logic overhead, and the advanced mechanisms used to overlap computation with memory transfers. As a result, two kernels with identical tile sizes may exhibit significantly different execution efficiencies due to these library‑specific optimizations—behavior that NeuSight’s theoretical modeling does not capture.

This issue is compounded by NeuSight’s reliance on PyTorch’s Linear and Batched MatMul (BMM) layers. For instance, PyTorch’s Linear layer often uses the TN transpose mode, where the first matrix is transposed before multiplication. In contrast,  torch.matmul or frameworks like ONNX and TensorFlow frequently use the NN transpose mode. In our experiments, different transpose mode can lead to different selections of library, algorithm, tilesize. 

NeuSight also does not account for advanced optimization techniques such as Split-K, reduction schemes, and swizzling \cite{nvidia2025cutlass}. These techniques, implemented within cuBLAS and CUTLASS, can significantly affect memory access patterns and kernel efficiency—even though they do not alter the FLOP count. Ignoring these factors can introduce measurable errors, especially in high-precision latency estimation tasks.

\textbf{Dataset Matching and Scalability Issues:} To determine the optimal tile size for a given input/output configuration, NeuSight relies on a precollected dataset. This dataset maps input shapes to tile sizes based on the assumption that a close domain around a predefined input shape yield the same MatMul configuration. While this approach works within a constrained domain, it introduces two major challenges:
\begin{itemize}
    \item \textbf{Matching Overhead}: Searching for the closest match in the dataset adds latency to the prediction process.
    \item \textbf{Generalization}: When input dimensions are outside the dataset's range, prediction accuracy drops significantly.
\end{itemize}
To address this, we examined NVIDIA’s documentations \cite{nvidia2025cublas12workspace} and source code from ONNX and PyTorch. We found that the cublasLtMatmulAlgoGetHeuristic() API can directly return the optimal configuration—including tile size, library, Split-K, reduction scheme, and swizzle—for any given input. This API internally queries GPU-specific information and must be executed on the target device, but it enables fast and accurate configuration collection. By using this API during NAS preprocessing, PM2Lat avoids the need for dataset matching and significantly reduces prediction time.

\begin{comment}
\textbf{Oversimplified Utility Layer Modeling:} Another limitation of NeuSight lies in its treatment of utility layers. It assumes that these layers with the same input and output sizes behave similarly, regardless of the operation type. However, our experiments reveal that this assumption does not hold. For example, binary operations (e.g., adding two tensors) and unary operations (e.g., adding a scalar to a tensor) differ significantly in memory access patterns and computational load. These differences are especially pronounced in memory-bound scenarios, where bandwidth becomes the dominant factor.

By generalizing utility layer behavior using two key metrics—the number of mathematical operations and the amount of memory accessed—PM2Lat captures these nuances more effectively. This abstraction allows for more accurate latency estimation across a wider range of utility layer types.
\end{comment}

\textbf{DNN-Based Prediction Overhead:} NeuSight and similar methods rely on DNN to predict kernel latency. While DNNs are powerful tools for modeling complex behaviors, they come with high computational costs. Training and evaluating these models can be time-consuming, especially when applied to millions of configurations during NAS.

In contrast, PM2Lat takes a more analytical approach. By studying kernel behavior and building lightweight regression models, it avoids the overhead of DNN training and inference. This makes PM2Lat faster, more interpretable, and easier to extend to new kernels and devices.

\subsection{PM2Lat: A Generalized and Lightweight Latency Prediction Framework.} \label{subsec.PM2Lat}
\textbf{MatMul Latency Prediction:} To build a latency prediction model that is both accurate and efficient, we begin by analyzing the behavior of MatMul kernels—one of the most common and compute-intensive operations in DL. MatMul has multiple implementations and configurations, making its performance characteristics complex to model. Rather than relying on the GPU to automatically select the best configuration, we manually specify kernel settings and analyze their behavior in isolation. This controlled approach allows us to understand how each configuration affects latency.

We follow a structured data collection strategy. Matrices are divided into tiles and assigned to thread blocks, which execute in waves—sequential batches of blocks processed by the GPU. Two key behaviors must be considered:
\begin{itemize}
    \item A thread block will execute fully even if its assigned tile is only partially filled.
    \item The final wave may contain fewer tiles than a full wave, but all blocks in that wave still execute in parallel.
\end{itemize}
%Please see appendix \ref{sec.APD.PartialCase} for a detail computation guideline for partial cases. 

To reduce variability, we collect data only from configurations that yield complete blocks and waves. We fix the tile size and number of waves, then vary the K dimension (Figure \ref{fig:TileGEMM})---which plays a critical role in performance----to observe its impact. Each kernel is executed at least 25 times with about 500ms as minimum total time of execution, and latency values are averaged after a warm-up period to ensure stability. The same strategy is also applied for partial MatMul cases.

\begin{figure}
    \centering
    \includegraphics[width=0.8\linewidth]{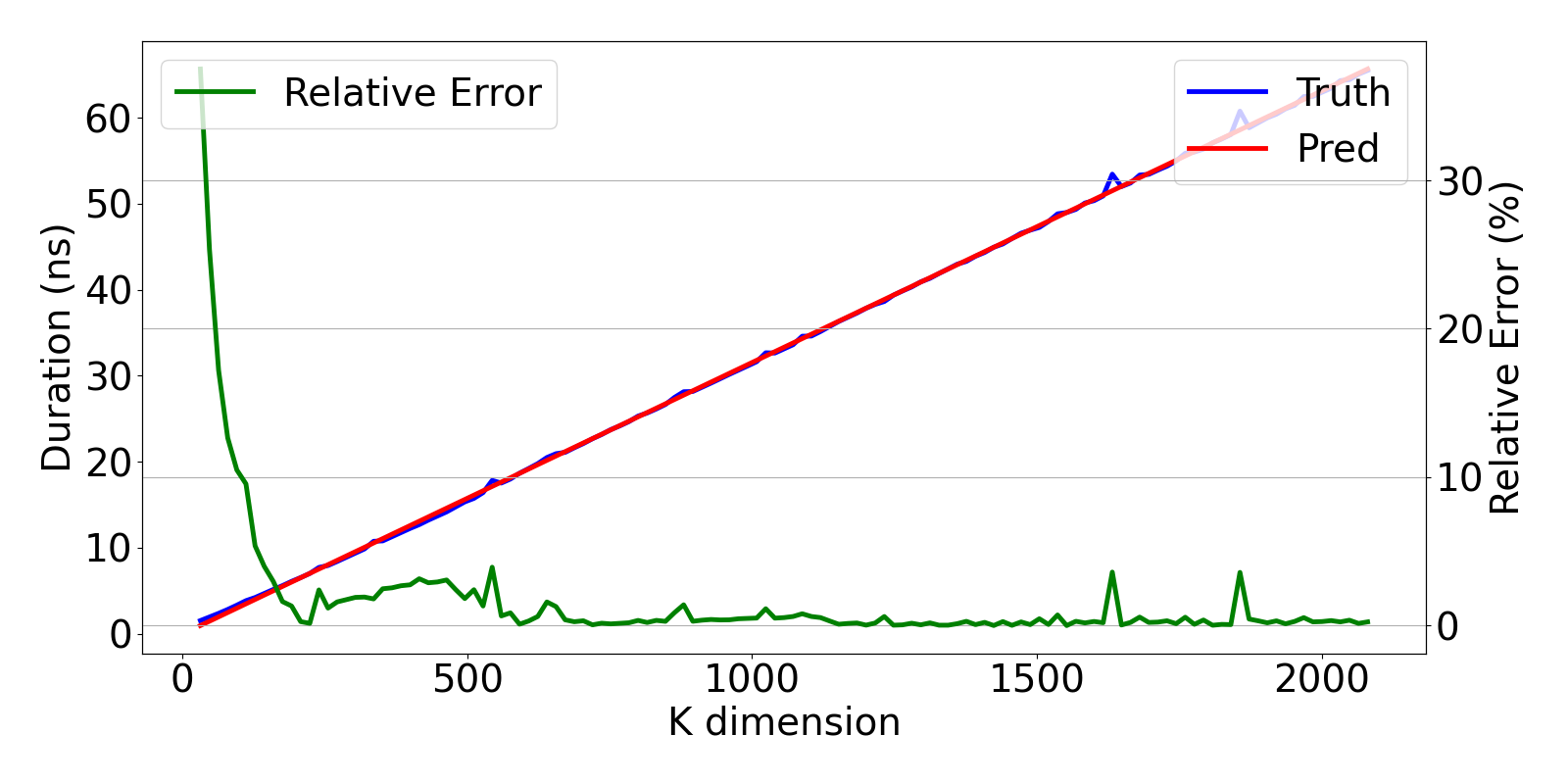}
    %\vspace{-0.3cm}
    \caption{Duration of kernel vs K when fixing number of waves and MatMul configuration.}
    \label{fig:DurvsC}
\end{figure}

Our experiments show that latency increases linearly with K under fixed settings following SIMT architecture with lockstep, but prediction errors grow at smaller K, as seen in Figure \ref{fig:DurvsC}. This is due to underutilization of GPU parallelism \cite{nvidia2023gpuperf} and the limitations of linear regression over wide input ranges. To address this, we analyze throughput---the number of operations completed per unit time---under controlled conditions, including fixed GPU frequency.

\begin{figure}
    \centering
    \includegraphics[width=0.8\linewidth]{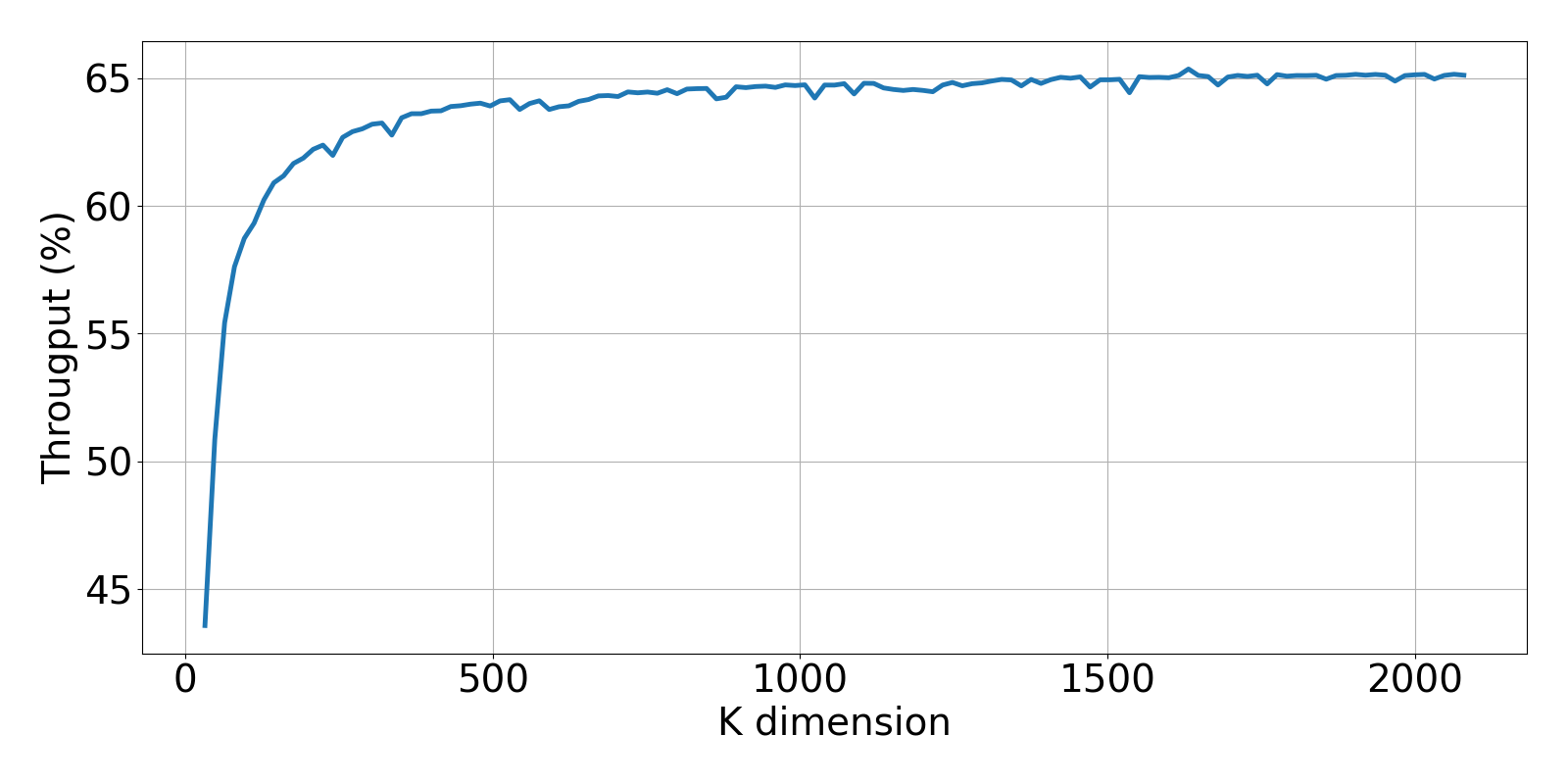}
    %\vspace{-0.3cm}
    \caption{Throughput of Kernel vs K when fixing frequency, number of wave and MatMul configuration.}
    \label{fig:ThrPutvsC}
\end{figure}

As shown in Figure \ref{fig:ThrPutvsC}, we find that throughput follows a rational trend ($y=(a\times x+b)/(c\times x +d)$) with respect to K. 
%{CP version here}
We also experimented with logarithmic approximations; however, they tended to fit the curve poorly. 

Instead of fitting a complex regression model, we collect throughput data at discrete powers-of-two values of K (e.g., 32, 64, 128, 256, ..., 8192). For new K values, we estimate latency using a simple interpolation formula:
\begin{equation}
    \text{newDur} = \text{orgDur} \times \frac{\text{newK}}{8192} \times \frac{\text{orgThrPut}}{\text{newThrPut}}
    \label{eq:DurInterpolation}
\end{equation}

%{CP version here}
Here, orgDur denotes the measured latency at $K=8192$, as beyond this point the throughput is unlikely to change further, and newThrPut is the interpolated throughput for the target K. The interpolation itself is computed using:
\begin{equation}
    \text{newThrPut} = \frac{K_{\text{new}} - K_1}{K_3 - K_1} \times (\text{ThrPut}_3 - \text{ThrPut}_1) + \text{ThrPut}_1
    \label{eq:InterpolationMatMul}
\end{equation}
where $K_{\text{new}}$ is the new K value, $K_1$ and $K_3$ are the closest under and above K values from the dataset, $\text{ThrPut}_1$ and $\text{ThrPut}_3$ is the corresponding throughput of $K_1$ and $K_3$, and $\text{newThrPut}$ is the target throughput for $K_{\text{new}}$.

\begin{figure*}[] % placement: t=top, h=here, etc.
  \centering
  \begin{subfigure}{0.32\textwidth}
    \centering
    \includegraphics[width=\linewidth]{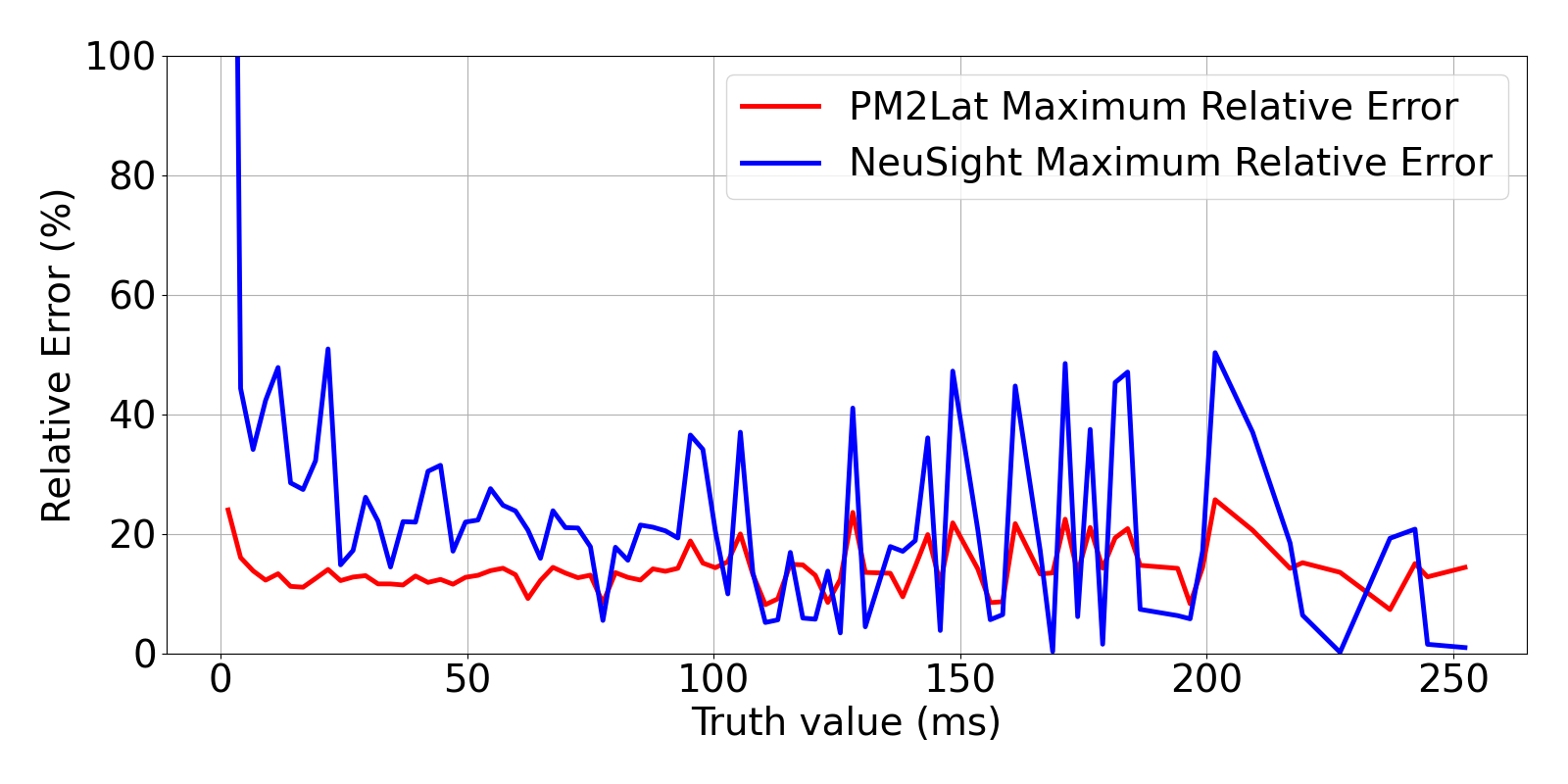}
    \caption{RTX3060M with FP32}
    \label{fig:first}
  \end{subfigure}\hfill
  \begin{subfigure}{0.32\textwidth}
    \centering
    \includegraphics[width=\linewidth]{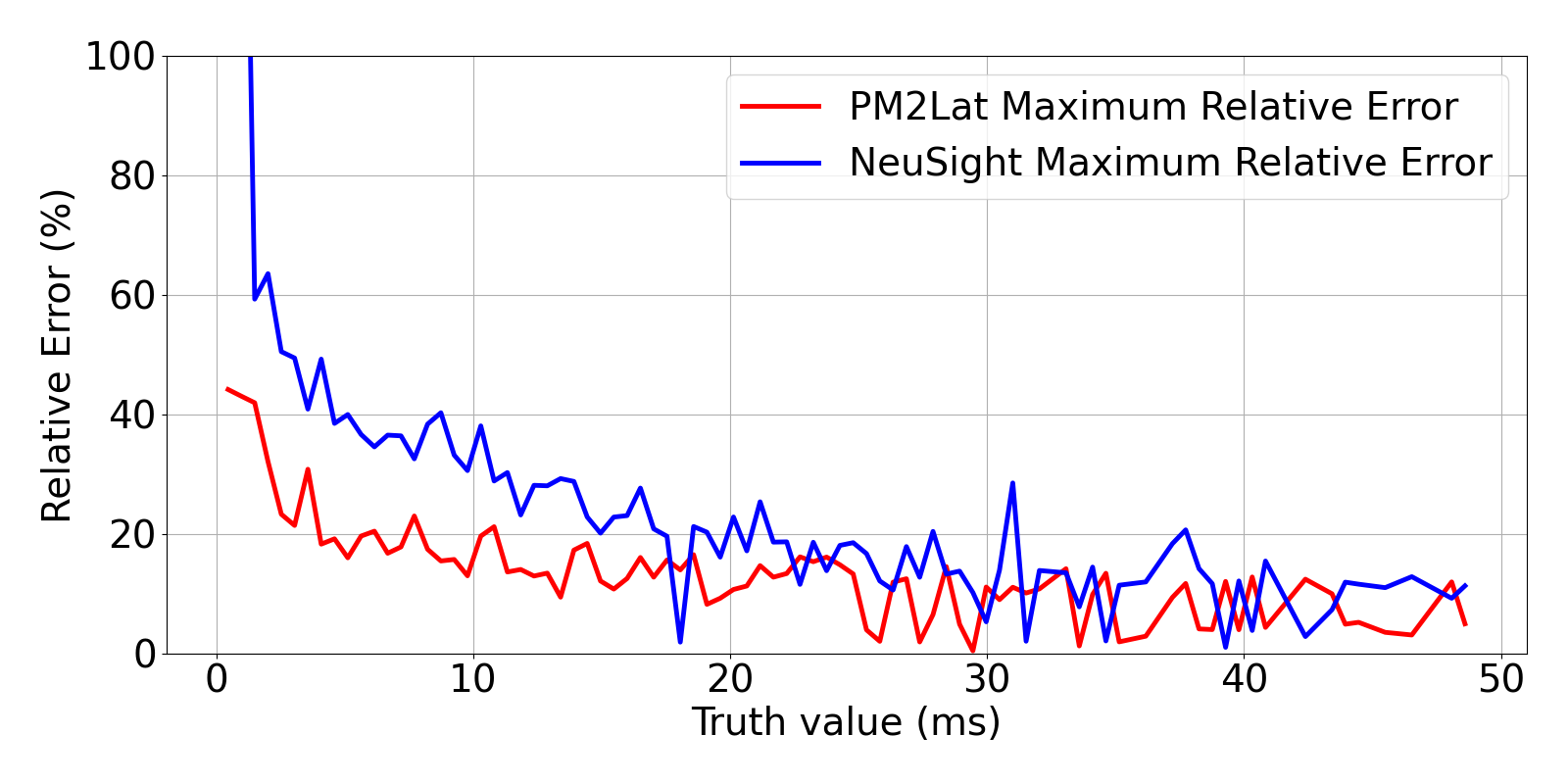}
    \caption{L4 with BF16}
    \label{fig:second}
  \end{subfigure}\hfill
  \begin{subfigure}{0.32\textwidth}
    \centering
    \includegraphics[width=\linewidth]{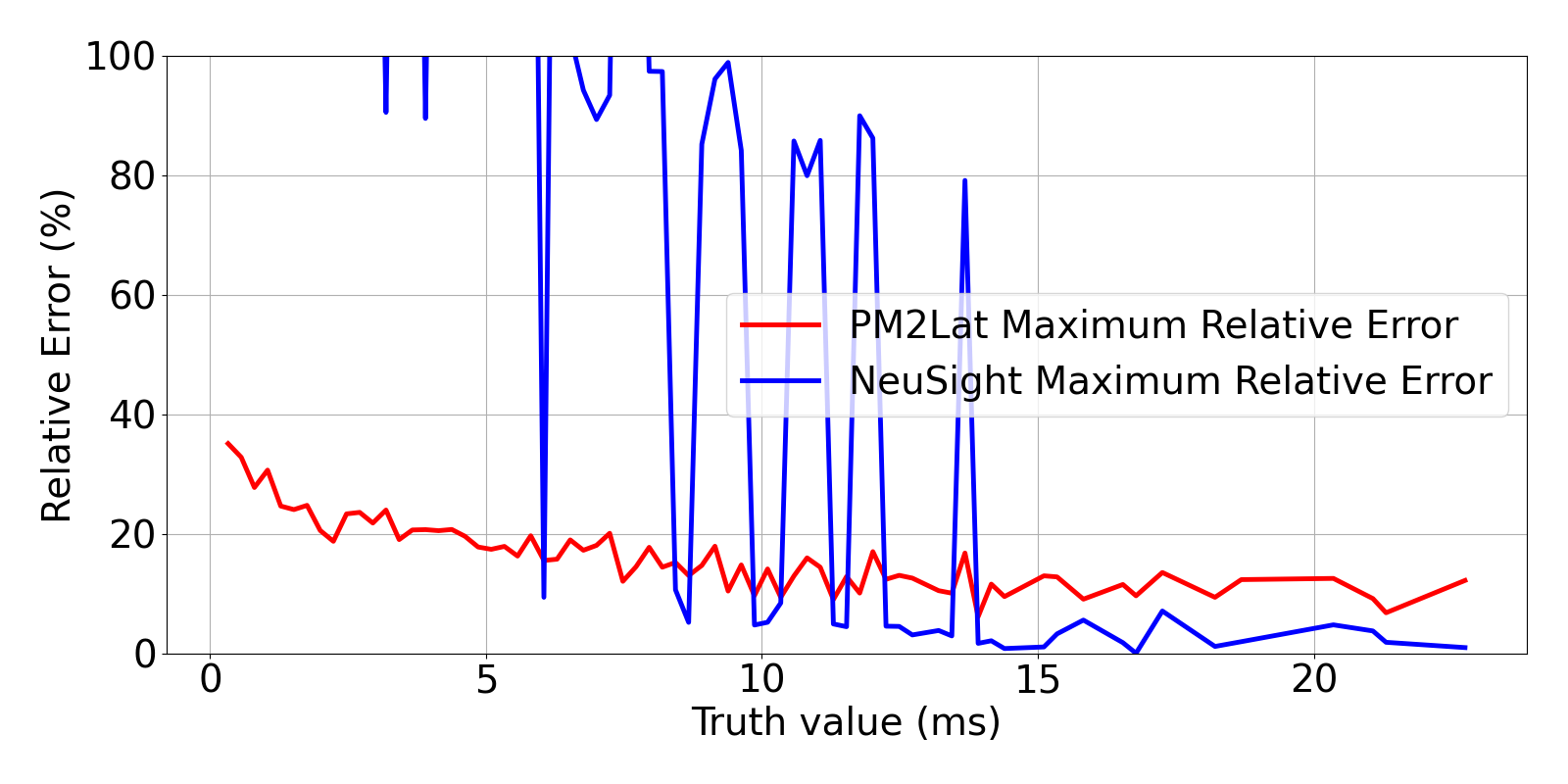}
    \caption{A100 with BF16}
    \label{fig:third}
  \end{subfigure}

  \caption{Maximum Relative Error of MatMul kernels.}
  \label{fig:MaxERRMatMul}
  \vspace{-0.6cm}
\end{figure*}

\begin{comment}
\begin{figure}[ht]
    \centering
    \includegraphics[width=0.75\linewidth]{Figs/MaxErr/ThanhLenovo_matmul_32F.png}
    \vspace{-0.3cm}
    \caption{Maximum Relative Error of MatMul kernels on RTX3060M (FP32).}
    \label{fig:MaxERRMatMulThanhLenovoFP32}
    \vspace{-0.5cm}
\end{figure}

\begin{figure}[ht]
    \centering
    \includegraphics[width=0.75\linewidth]{Figs/MaxErr/ColabL4_matmul_16BF.png}
    \vspace{-0.3cm}
    \caption{Maximum Relative Error of MatMul kernels on L4 (BF16).}
    \label{fig:MaxERRMatMulColabL4BF16}
    \vspace{-0.5cm}
\end{figure}

\begin{figure}[ht]
    \centering
    \includegraphics[width=0.75\linewidth]{Figs/MaxErr/ColabA100_matmul_16BF.png}
    \vspace{-0.3cm}
    \caption{Maximum Relative Error of MatMul kernels on A100 (BF16).}
    \label{fig:MaxERRMatMulColabA100BF16}
\end{figure}
\end{comment}

This method allows us to estimate latency for any K value with high accuracy and minimal overhead. Please note that this step will be repeated for each MatMul configuration. Hence, if we know the optimal MatMul configuration for the target input parameters, we can use this method to estimate its latency.

All latency and throughput measurements are collected using NVIDIA CUPTI, which provides low-overhead access to GPU hardware counters.

\textbf{Utility Layer Latency Prediction:} Utility layers such as GeLU, ReLU, and Softmax are less compute-intensive but highly sensitive to memory bandwidth. Their latency depends more on how data is accessed than on how much computation is performed. Since cache bandwidths (L1, L2) are not directly observable, we use proxy metrics: amount of memory accessed, and number of executed instructions (FLOPs, integer ops, memory loads/stores).
%\begin{itemize}
%    \item Amount of memory accessed
%    \item Number of executed instructions (FLOPs, integer ops, memory loads/stores)
%\end{itemize}

%These metrics are collected using NVIDIA Nsight Compute (NCU), and latency is measured using CUPTI. We then apply linear regression to model the relationship between these metrics and latency. 
%With this approach, we can know the actual information of the operation, reducing the redundant between theoretical computation and the real one, as well as ignoring the differences in behavior across layer types. 
%With this approach, we can obtain the actual operational information, reducing the redundancy between theoretical and real computations, while also ignoring behavioral differences across layer types.
%For new inputs or devices, we scale the metrics and apply the coefficients to estimate latency.

%{CP version here}
These metrics are collected using NVIDIA Nsight Compute (NCU), and latency is measured using CUPTI. We then apply linear regression to model the relationship between these metrics and latency. With this approach, we base our model entirely on actual implementation‑level behavior---capturing effects such as indexing overhead, memory transactions, and control flow---rather than relying on theoretical formulas or assumptions about how different layer types behave. This allows us to avoid hand‑crafted per‑layer analytical models, which often require strong assumptions and do not reflect real kernel execution. For new inputs or devices, we scale the measured metrics and apply the learned coefficients to estimate latency. This scaling step 
%does not introduce a theoretical model; instead, it 
normalizes empirically collected, implementation‑specific metrics so they can generalize across devices whose memory‑bound kernels typically show similar performance characteristics.

\begin{comment}
\paragraph{\textbf{Generalization to Other Kernels.}} Our experiments show that linear regression works well when kernel configurations are clearly defined. For example, each cuBLAS/CUTLASS MatMul configuration is treated as a distinct kernel. This insight allows us to extend PM2Lat to other compute-intensive kernels, such as Triton, Fused Multi-Head Attention, Convolution, and custom CUDA kernels.

However, care must be taken when kernel dimensions are not fixed. If any dimension—like K in MatMul—is variable or does not contribute to the block scheduling step, performance can vary significantly. In such cases, modeling the relationship between \textbf{throughput} and \textbf{the unfixed dimension} improves prediction accuracy. Section 4.3 shows how this strategy applies to Triton, Flash Attention, and Cutlass Attention kernels.

Finally, to estimate the latency of an entire DNN model, we simply sum the predicted latencies of all layers, assuming sequential execution. This mirrors the behavior of CUDA kernels on GPUs, where layers are sequentially executed.
\end{comment}

\section{Evaluation and Discussion} \label{sec.Evaluation}

To assess the effectiveness of PM2Lat, we conduct a comprehensive evaluation at both the layer and model levels. Our goal is to measure how accurately PM2Lat predicts latency across different types of layers and hardware platforms, and how it compares to NeuSight—the current state-of-the-art method. We do not include Habitat in our comparison, as NeuSight has already demonstrated superior performance to Habitat, making NeuSight the more relevant baseline for evaluating PM2Lat.

We begin with evaluating latency prediction at the individual layer level, focusing on common operations such as Batched Matrix Multiplication (BMM), Linear, MatMul, and utility layers. These layers are \textit{sampled} using randomly generated input and output sizes to ensure diversity and coverage. We then extend the evaluation to full DNN models, using \textit{Transformer architectures} as the representative examples. Finally, we demonstrate how PM2Lat can be adapted to other GPU kernels, highlighting its generalizability.

Our experiments span five NVIDIA devices—RTX 3060M, T4, L4, A100 and RTX5070—covering a range from mobile-grade to server-grade architectures. 
This diversity allows us to test PM2Lat under varied computational capabilities and thermal conditions. The detail information of these devices are specified in Table \ref{tab:GPUInfo}. Furthermore, we evaluate performance under both FP32 and BF16 precision settings, and re-collect and re-train NeuSight models to align with each data type. It should be noted that T4 does not support the BF16 data type. 
\begin{table}[ht]
    \centering
    \caption{Specification of tested GPUs.}
    \label{tab:GPUInfo}
    %\vspace{-8pt}
    \small
    \begin{tabular}{|c|c|c|c|c|c|}
        \hline
         & \cellcolor{gray!32}3060M &  \cellcolor{gray!32}T4 &  \cellcolor{gray!32}L4& \cellcolor{gray!32}A100& \cellcolor{gray!32}5070\\ 
         \hline
         %Arch& Ampere& Tesla& Ampere&Ampere\\
         %\hline
         %Device& Laptop& Server& Server&Server\\
         %\hline
         Max Freq (GHz)& 2.090&1.590& 2.040&1.410&3.090\\
         \hline
         %Base Freq& \multirow{2}{*}{0.900}&\multirow{2}{*}{0.595}&\multirow{2}{*}{0.795}&\multirow{2}{*}{0.765}\\
         %(MHz)&&&&\\
         %\hline
         FP32 (TFLOPs)& 16.05&8.141&30.29&19.49&37.97\\
         \hline
         BF16 (TFLOPs)& 32.10&-&121.16&311.87&75.94\\
         \hline
         DRAM BW (GB/s)& 336&320&300&1560&672\\
         \hline
         MEM (GB)& 6& 16&24&40&12\\
         \hline
         L2 (MB)& 3& 4&48&40&48\\
         \hline
         SM Count & 30& 40&58&108&48\\
         \hline
         No.CUDA.Cores & 3840& 2560&7242&6912&6144\\ 
         \hline
         Power (W) & 130 & 70 & 70 & 400 & 250 \\
         \hline
    \end{tabular}
    %\vspace{-10pt}
\end{table}

\subsection{Per-Layer Latency Prediction}\label{subsec.Eval.PerLayer}

\begin{table}[h!]
\centering
\caption{Average relative error (\%) comparison between PM2Lat (PL) and NeuSight (NS) across different layer types.}
\label{table:LayerWiseComp}
\begin{tabular}{|c|l|l|c|c|c|c|c|}
\hline
\textbf{DType} & \textbf{Layer} & \textbf{} &\textbf{3060M} & \textbf{T4} & \textbf{L4} & \textbf{A100} & \textbf{5070}\\
\hline
&BMM & NS & 7.8& 27.0& \textbf{\color{red}11.1}& \textbf{\color{red}1.8}& 9.1 \\
&    & PL   & \textbf{\color{red}2.3}& \textbf{\color{red}18.1}& 21.4& 2.0& \textbf{\color{red}8.2}\\ \cline{2-8}
&MM  & NS &        11.0& 13.1& 12.0& 4.0& 7.8\\
&    & PL   &        \textbf{\color{red}6.0}& \textbf{\color{red}12.1}& \textbf{\color{red}8.8}& \textbf{\color{red}3.1}&\textbf{\color{red}5.1}\\\cline{2-8}
Float32&Linear & NS &    7.2&         25.3&         12.6&         6.0& 21.5\\
&       & PL   &    \textbf{\color{red}4.0}&         \textbf{\color{red}16.7}&         \textbf{\color{red}7.1}&         \textbf{\color{red}3.1}&\textbf{\color{red}7.2}\\\cline{2-8}
&SoftMax & NS & 29.9& \textbf{\color{red}18.1}& \textbf{\color{red}23.9}& 27.8&\textbf{\color{red}7.4}\\
&        & PL   & \textbf{\color{red}11.9}& 20.2& 44.9& \textbf{\color{red}8.2}&8.6\\\cline{2-8}
&Vector & NS & 38.0& 37.0& 50.5& 35.4&36.9\\
&       & PL   & \textbf{\color{red}0.8}& \textbf{\color{red}1.7}& \textbf{\color{red}14.1}& \textbf{\color{red}1.4}&\textbf{\color{red}5.4}\\\hline

&BMM & NS & 17.7& -& \textbf{\color{red}27.0}& 38.5& 28.1\\
&    & PL   & \textbf{\color{red}4.5}& -& 32.6& \textbf{\color{red}9.3}&\textbf{\color{red}5.4}\\ \cline{2-8}
&MM  & NS &        15.1& -& 24.5& 62.3&50.7\\
&    & PL   &        \textbf{\color{red}6.6}& -& \textbf{\color{red}8.5}& \textbf{\color{red}12.9}&\textbf{\color{red}4.4}\\\cline{2-8}
BFloat16&Linear & NS &    18.9&         -&         28.2&         70.5&46.6\\
&       & PL   &    \textbf{\color{red}6.4}&         -&         \textbf{\color{red}8.4}&         \textbf{\color{red}10.3}&\textbf{\color{red}5.9}\\\cline{2-8}
&SoftMax & NS & 43.5& -& 127.1& 19.9&43.2\\
&        & PL   & \textbf{\color{red}9.2}& -& \textbf{\color{red}31.7}& \textbf{\color{red}7.8}& \textbf{\color{red}19.9}\\\cline{2-8}
&Vector & NS & 72.2& -& 115.3& 14.2&82.9\\
&       & PL   & \textbf{\color{red}3.3}& -& \textbf{\color{red}19.4}& \textbf{\color{red}2.6}&\textbf{\color{red}5.3}\\\hline
\end{tabular}
\end{table}

We collect 1000 different latency samples for each layer type corresponding to each datatype. For BMM kernels, dimensions are capped at 1024. For MatMul and Linear layers, M and N dimensions go up to 8192, while K is limited to 20000. Utility layers are tested with batch sizes and input features up to 16384.

Table \ref{table:LayerWiseComp} presents the average relative error across different layer types, data types and devices. PM2Lat outperforms NeuSight in most cases, with particularly huge gap on the BF16 data type. For example, with BF16 on A100, PM2Lat achieves a 10.3\% error on Linear layers compared to NeuSight’s 70.5\%. The large gap observed when switching to BF16 arises because, in FP32, NVIDIA libraries offer about 13 combinations of algorithms and tile sizes for computing MatMul layers, whereas BF16 provides nearly 100. The higher number of combinations increases the contribution of unseen factors such as memory access patterns as mentioned in Section~\ref{sec.ProposedMethod}. This count does not even include additional factors such as splitK, swizzle, stage and reduction schemes. Consequently, when relying solely on FLOPs and wave counts---without accounting for performance differences among these combinations---NeuSight tends to exhibit lower accuracy as the performance disparity between algorithms increases. Moreover, NeuSight’s “sieve” strategy---where data samples are proportionally distributed across the input domain---can improve accuracy for inputs that closely resemble training data. However, this approach struggles when input dimensions deviate significantly from the training distribution, leading to higher marginal errors.%, as demonstrated in Figure \ref{fig:MaxERRMatMulColabA100BF16}. In contrast, PM2Lat differentiates performance across kernels, ensuring stable error rates regardless of data type.}. %This consistency stems from PM2Lat’s design, which avoids overfitting to specific hardware or configurations.  In contrast, PM2Lat’s interpolation-based method generalizes better across unseen configurations, making it more robust in practical deployment scenarios.

\begin{figure}
    \centering    
    \includegraphics[width=0.8\linewidth]{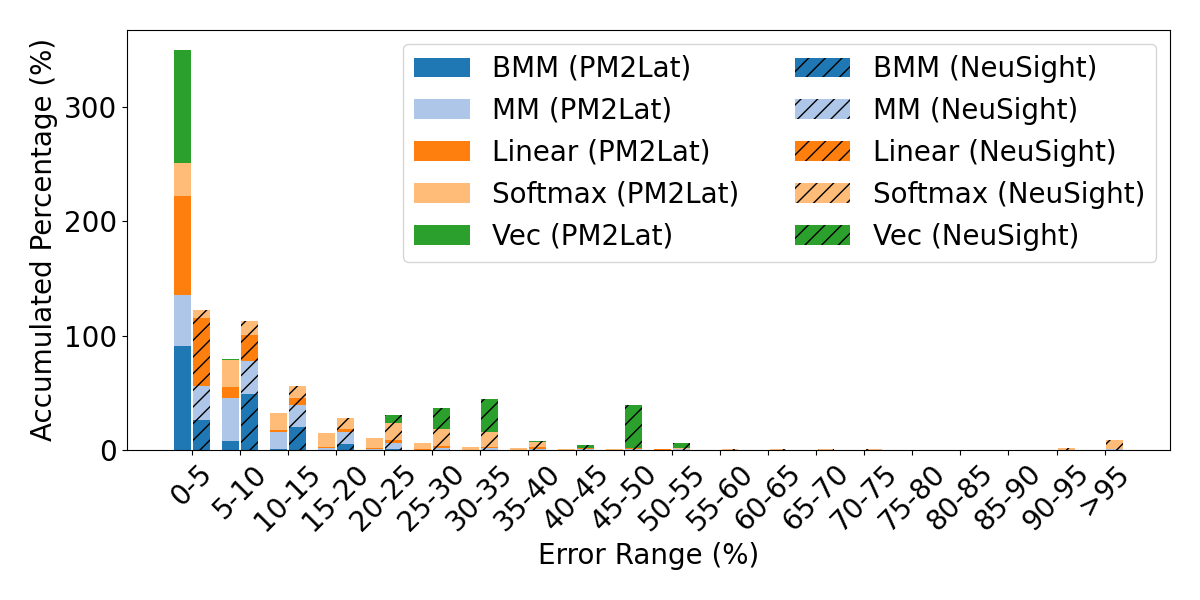}
    \vspace{-0.3cm}
    \caption{Error distribution on RTX3060M (FP32).}
    \label{fig:ErrDist3060FP32}
    \vspace{-0.3cm}
\end{figure}

\begin{figure}
    \centering
    \includegraphics[width=0.8\linewidth]{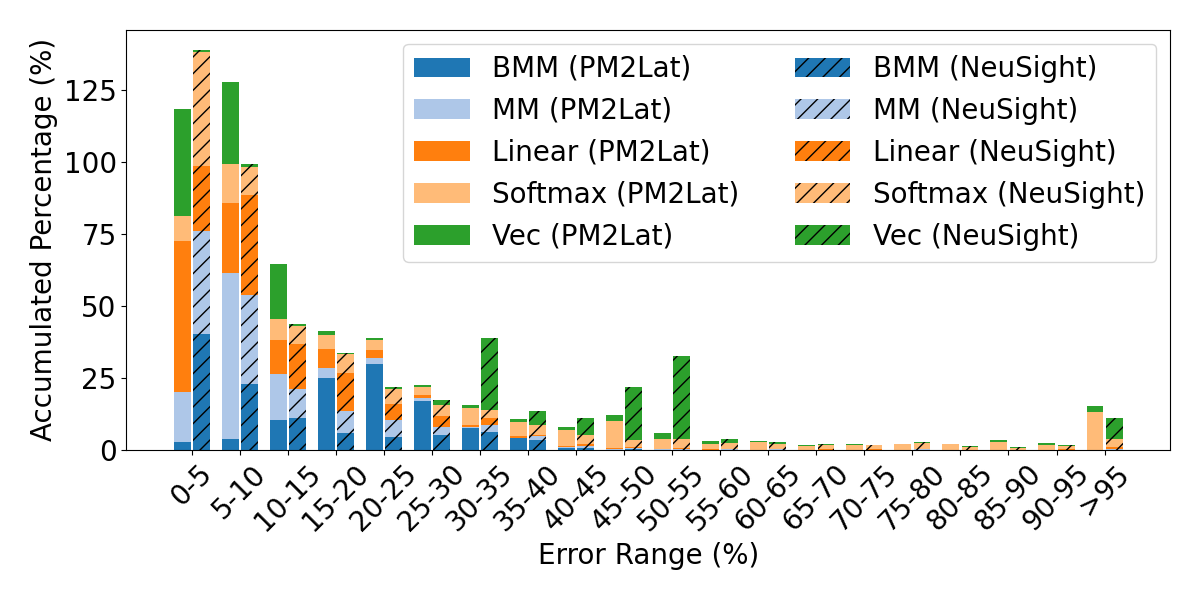}
    \vspace{-0.3cm}
    \caption{Error distribution on RTX5070 (FP32).}
    \label{fig:ErrDist5070FP32}
    %\vspace{-0.6cm}
\end{figure}

%Figures \ref{fig:MaxERRMatMulThanhLenovoFP32}, \ref{fig:MaxERRMatMulColabL4BF16} and \ref{fig:MaxERRMatMulColabA100BF16} illustrate the worst-case prediction errors for both PM2Lat and NeuSight. 
Figure \ref{fig:MaxERRMatMul} illustrates the worst-case prediction errors for both PM2Lat and NeuSight. 
To highlight substantial deviations, the input domain is divided into 100 bins, and only the maximum error in each bin is plotted. The results show that NeuSight consistently exhibits higher worst-case errors, especially with BF16 data type on A100, whereas PM2Lat maintains stable performance regardless of device, data type, or layer type. This suggests that NeuSight has limitations in generalization, raising concerns about its ability to accurately predict latency on unseen hardware configurations. In contrast, PM2Lat’s kernel-aware approach enables robust predictions across diverse environments, making it more suitable for heterogeneous deployment scenarios. 
%{CP version here}
Interestingly, NeuSight’s performance on the L4 GPU with BF16 is quite good, yielding accuracy levels close to those of our method—an outcome that contrasts with its performance on the A100. This behavior is influenced by the inherent bias of the DNN model toward patterns present in its training dataset, as discussed in Section \ref{subsec.Eval.ModelWise}. However, the metric shown here reflects only the maximum relative error within each data bin, representing the worst‑case scenario. To understand NeuSight’s overall behavior, we must also examine the average error and the full distribution across all datapoints.

\begin{figure}
    \centering
    \includegraphics[width=0.8\linewidth]{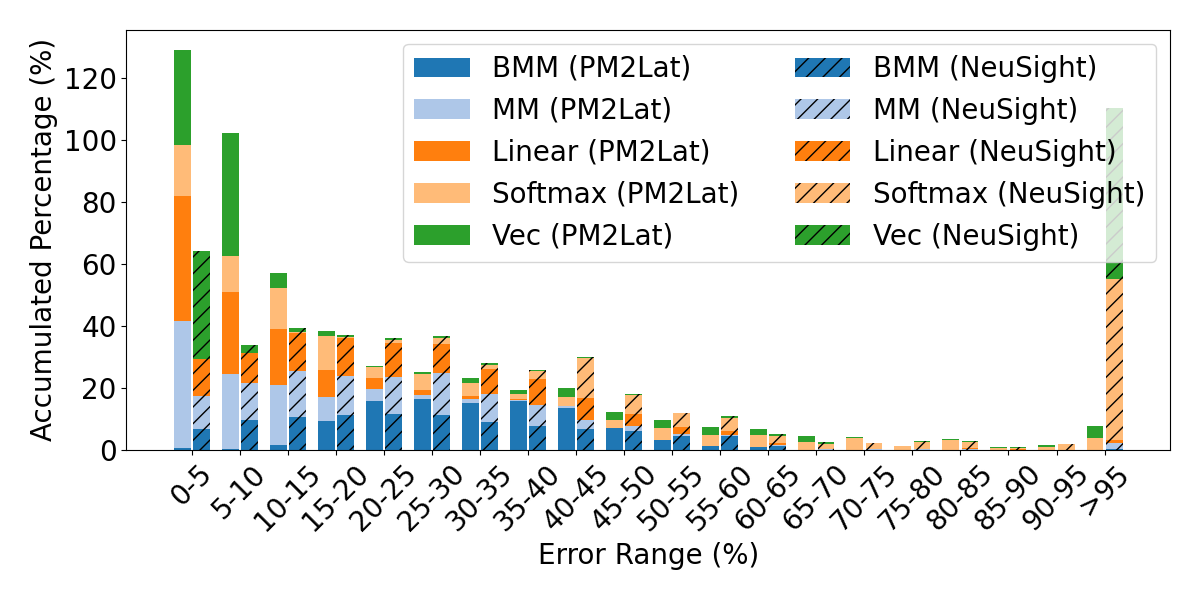}
    \vspace{-0.3cm}
    \caption{Error distribution on L4 (BF16).}
    \label{fig:ErrDistL4BF16}
    \vspace{-0.3cm}
\end{figure}

\begin{figure}
    \centering
    \includegraphics[width=0.8\linewidth]{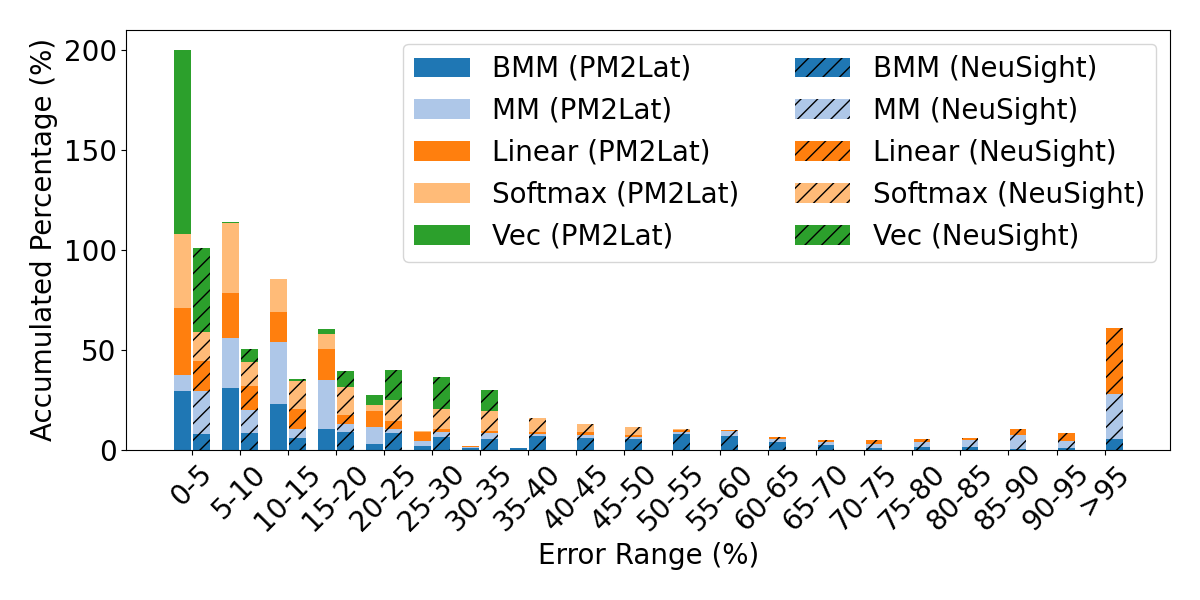}
    \vspace{-0.3cm}
    \caption{Error distribution on A100 (BF16).}
    \label{fig:ErrDistA100BF16}
    %\vspace{-0.4cm}
\end{figure}

Figures \ref{fig:ErrDist3060FP32}, \ref{fig:ErrDist5070FP32}, \ref{fig:ErrDistL4BF16}, and  \ref{fig:ErrDistA100BF16} visualize how the prediction error is distributed. For FP32, both methods exhibit a promising trend, with errors concentrated at lower values. Specifically, PM2Lat achieves a higher proportion of predictions with error rates below 15\%. When switching to BF16, PM2Lat retains these characteristics, whereas NeuSight shows a significant increase in predictions with error rates exceeding 95\%, indicating poor robustness under datatype changes.
Furthermore, when training NeuSight with BF16 on a single device, it achieves only marginal improvements, further confirming the limitations of this state-of-the-art method.

%{CP version here}
Thermal behavior also plays a critical role in latency prediction accuracy, especially on devices with limited cooling capabilities. For instance, GPUs like the T4 and L4 rely on passive cooling mechanisms, making them more vulnerable to thermal throttling under sustained workloads. NeuSight’s profiling process, which involves collecting large datasets at high operational frequencies, tends to generate more heat. This allows it to not only capture thermal characteristics more effectively---especially on strong workload like BMM---but also amplifies performance disparities across devices---particularly when comparing actively cooled versus passively cooled systems, as NeuSight relies solely on several specific GPU information like theoretical peak FLOP per second rather than actual behavior of a device.

In contrast, PM2Lat uses a smaller number of samples and primarily collects throughput data at lower GPU frequencies. As a result, it does not capture the thermal behavior of devices under stress, leading to an average relative error of 32.6\%, compared to NeuSight’s 27.0\%. While this design choice contributes to PM2Lat’s efficiency and stability, it limits its ability to model temperature-induced latency variations. Accurately analyzing thermal effects would require more complex experimental setups and longer profiling durations, which are beyond the scope of this work. Hence, we leave this challenge as an open direction for future research.

Overall, PM2Lat demonstrates robust and consistent latency prediction across a wide range of layer types and hardware. However, this subsection focuses exclusively on per-layer accuracy. In practice, the overall model accuracy is typically more critical, as errors across individual layers accumulate and impact end-to-end performance. Therefore, subsection \ref{subsec.Eval.ModelWise} presents our evaluation of model-wise accuracy.
%Its lightweight design, memory-aware modeling, and interpolation based strategy allow it to outperform NeuSight in most scenarios—especially on utility layers. These results validate PM2Lat’s practicality for real-world deployment and large-scale NAS, where speed, accuracy, and generalizability are essential.

\subsection{Model Latency Prediction} \label{subsec.Eval.ModelWise}       

In this subsection, we conduct end-to-end latency prediction experiment using several Transformer architectures, including GPT-2, FLAN-T5, Qwen 3 and DeepSeek R1 \cite{radford2019language,chung2024scaling,qwen3technicalreport,deepseekai2025deepseekr1incentivizingreasoningcapability}, shown in Table \ref{tab:model_comparison}. 
%These models are representative of modern workloads, including a mix of compute- and memory-intensive layers. 
GPT-2 and FLAN-T5 pretrained models come originally with FP32 whereas BF16 is applied to Qwen 3 and DeepSeek R1. To ensure consistency and reproducibility, we also employ ONNX beside PyTorch frameworks for model execution, where GPT-2 and FLAN-T5 will be converted to ONNX. 
%As previously noted, these frameworks differ in their matrix transpose conventions for dense layers: ONNX adopts an NN layout, whereas PyTorch uses TN. This discrepancy leads to variations in tile sizes and the algorithms selected during execution.
Each model undergoes a warm-up phase with five repetitions, followed by 25 latency measurements to compute the average runtime. We focus exclusively on the inference phase, as the backward pass in training typically reuses MatMul operations and does not introduce additional kernel types, as mentioned in NeuSight. Importantly, all model information will first be collected on the A100 GPU and then systematically mapped to the corresponding devices.

\begin{table}[t]
\centering
\caption{Overview of Selected Transformer Models}
\begin{tabular}{|l|c|c|c|c|}
\hline
\textbf{Model} & \textbf{Year} & \textbf{Params} & Framework & Datatype\\
\hline
%BERT (Base) & 2018 & 110M \\
%\hline
%BART (Base) & 2020 & 139M & Encoder-Decoder & Combines BERT-style encoder with GPT-style decoder; excels in text generation and summarization. \\
%\hline
%DistilBERT & 2019 & 66M & Encoder-only & Compressed version of BERT with ~97\% of its performance; optimized for speed and efficiency. \\
%\hline
%OPT (350M) & 2022 & 350M & Decoder-only & Open-source alternative to GPT; designed for efficient training and reproducibility. \\
%\hline
GPT-2 (Large) & 2019 & 774M& ONNX & FP32 \\
\hline
FLAN-T5 (Base) & 2023 & 250M & ONNX & FP32 \\
\hline
Qwen 3 - 0.6B & 2025 & 0.6B& Pytorch & BF16\\
\hline
Qwen 3 - 4B & 2025 & 4B & Pytorch & BF16\\
\hline
DeepSeek R1 - 7B & 2025 & 7B & Pytorch & BF16\\
\hline
DeepSeek R1 - 14B & 2025 & 14B & Pytorch & BF16\\
\hline
\end{tabular}
\label{tab:model_comparison}
\end{table}

Table \ref{tab:ModelCompare} and \ref{tab:ModelCompareCont} presents a comparison of accuracy between PM2Lat and NeuSight across different models, batch sizes, and hardware. 
For the FP32 models (GPT2 and FLAN-T5), both PM2Lat and NeuSight perform reasonably well at smaller batch sizes, but clear differences emerge as the workload scales. PM2Lat maintains relatively low error rates, typically under 10\%, even for larger batches.
In most of the cases on 3060M, T4 and L4 with FP32 datatype, PM2Lat totally outperforms NeuSight, typically more than 10\%. For example, GPT2 on T4 shows the error rate of 9.1\% for PM2Lat, which is a much lower value compared to 25.5\% for NeuSight at BS=32.
On A100 and 5070, PM2Lat performs comparably to NeuSight, particularly at higher batch sizes. This is expected, as the A100’s superior memory bandwidth and compute capacity reduce the impact of kernel-level inefficiencies. %However, even in this favorable setting, PM2Lat maintains competitive accuracy while offering significantly faster prediction times.

\begin{table*}[h]
    \centering
    \caption{Model-Wise Comparison of accuracy between PM2Lat (PL) and NeuSight (NS).}
    \label{tab:ModelCompare}
    \begin{tabular}{|p{0.6cm}|c|ccc|ccc|ccc|ccc|ccc|}
    \hline
        & & \multicolumn{3}{|c|}{3060M} & \multicolumn{3}{|c|}{T4} & \multicolumn{3}{|c|}{L4} & \multicolumn{3}{|c|}{A100}& \multicolumn{3}{|c|}{5070} \\
        Model& BS&MeanT&PL&NS&MeanT&PL&NS&MeanT&PL&NS&MeanT&PL&NS&MeanT&PL&NS\\
        &&(ms)&(\%)&(\%)&(ms)&(\%)&(\%)&(ms)&(\%)&(\%)&(ms)&(\%)&(\%)&(ms)&(\%)&(\%)\\ \hline

        & 1 & 32 & \textbf{\color{red}-2.5} & +16.5 & 45 & \textbf{\color{red}+10.0} & +27.6 & 19 & \textbf{\color{red}-6.6} & +47.1 & 16 & \textbf{\color{red}-11.1} & +27.4 & 12 & -25 & \textbf{\color{red}+2.3}\\

        & 8 & 232 & \textbf{\color{red}-3.5} & +6.4 & 343 & \textbf{\color{red}+11.6} & +23.4 & 188 & \textbf{\color{red}+1.8} & +21.1 & 85 & \textbf{\color{red}-0.1} & +7.6 & 99 & -5.5 & \textbf{\color{red}-3.5}\\

        GPT2 & 16 & 458 & \textbf{\color{red}-2.9} & +14.4 & 661 & \textbf{\color{red}+13.5} & +24.4 & 396 & \textbf{\color{red}+2.7} & +14.2 & 165 & \textbf{\color{red}-1.4} & +4.5 & 198 & \textbf{\color{red}-2.7} & +3.7\\

        & 32 & -& -& -& 1380 & \textbf{\color{red}+9.1} & +25.5 & 832 & \textbf{\color{red}+0.3} & +12.6 & 322 & \textbf{\color{red}+0.0} & +2.1 & 418 & -4.6 & \textbf{\color{red}-2.4} \\

        & 64 & -& -& -& -& -& -& -& -& -& 947 & \textbf{\color{red}+1.7} & +2.2 &-&-&-\\ \hline

        %%%%%%%%%%%%%%%%%%%%%%%%%%%%%%%%%%%%%%%%%%%%%%%%%%%%%%%%%%%%%%%%%%%%%%

        & 1 & 54 & \textbf{\color{red}+3.6} & +31.9 & 86 & \textbf{\color{red}+3.2} & +18.7 & 39 & \textbf{\color{red}-8.8} & +25.2 & 27 & \textbf{\color{red}+7.5} & +54.9 & 21 & -19.5 & \textbf{\color{red}-0.5}\\

        & 8 & 387 & \textbf{\color{red}+7.1} & +17.4 & 645 & \textbf{\color{red}+3.8} & +5.9 & 362 & \textbf{\color{red}+4.2} & +9.9 & 153 & \textbf{\color{red}-1.5} & +10.6 & 156 & +10.0 & \textbf{\color{red}+6.1}\\

        FLAN & 16 & 794 & \textbf{\color{red}+7.8} & +31.1 & 1249 & \textbf{\color{red}+7.0} & +10.9 & 754 & \textbf{\color{red}+1.3} & +8.3 & 295 & \textbf{\color{red}-1.1} & +7.9 & 330 & \textbf{\color{red}+7.3} & 11.7\\

        T5& 32 & 1569 & \textbf{\color{red}+6.9} & +31.9 & 2556 & \textbf{\color{red}+5.2} & +21.7 & 1459 & \textbf{\color{red}+3.5} & +11.7 & 572 & \textbf{\color{red}+0.7} & +3.2 & 679 & \textbf{\color{red}+5.4} & +7.1\\

        & 64 & -& -& -& 5105 & \textbf{\color{red}+5.3} & +21.1 & 2916 & \textbf{\color{red}+5.1} & +11.3 & 1126 & +2.2 & \textbf{\color{red}+1.6} & 1358 & \textbf{\color{red}+5.8} & +8.9\\ \hline

        %%%%%%%%%%%%%%%%%%%%%%%%%%%%%%%%%%%%%%%%%%%%%%%%%%%%%%%%%%%%%%%%%%%%%%

        & 1 & 58 & \textbf{\color{red}+15.4}& +63.3 & -& - & - & 49& \textbf{\color{red}-49.1}& +80.7& 47& -51.4& \textbf{\color{red}+3.5}& 23& \textbf{\color{red}-26.9}& +90.4\\
        Qwen& 8& 422& \textbf{\color{red}+1.1}& +53.9& - & - & - & 310& \textbf{\color{red}+16.8}& +76.8& 91& \textbf{\color{red}+9.1}& +114.3& 184& \textbf{\color{red}+9.3}& +90.0\\
        3& 16& -& -& -& - & - & - & 687& \textbf{\color{red}+13.0}& +61.6& 174& \textbf{\color{red}+7.3}& +104.7& 389& \textbf{\color{red}+7.2}& +85.1\\
        0.6B& 32& -& -& -& - & - & - & 1489& \textbf{\color{red}+7.6}& +48.0& 332& \textbf{\color{red}+8.7}& +90.8& -& -& -\\
        & 64& -& -& -& - & - & - & -& -& -& 652& \textbf{\color{red}+8.8}& +83.6& -& -& -\\ \hline

        %%%%%%%%%%%%%%%%%%%%%%%%%%%%%%%%%%%%%%%%%%%%%%%%%%%%%%%%%%%%%%%%%%%%%%

        & 1 & -& -& -& - & - & - & 135& \textbf{\color{red}-7.0}& +83.8& 61& \textbf{\color{red}-13.0}& +83.4& 95& \textbf{\color{red}-5.4}& +56.5\\
        Qwen& 8& -& -& -& - & - & - & 1235& \textbf{\color{red}+8.7}& +49.1& 298& \textbf{\color{red}+5.5}& +134.0& -& -& -\\
        3 4B& 16& -& -& -& - & - & - & 2545& \textbf{\color{red}+9.4}& +48.1& 567& \textbf{\color{red}+8.3}& +138.7& -& -& -\\
        & 32& -& -& -& - & - & - & -& -& -& 1115& \textbf{\color{red}+8.8}& +138.7& -& -& -\\ \hline
        
    \end{tabular}\\
    MeanT: Mean Time of real executions. "-" is out-of-memory (OOM)
    %An extended version with more models can be found in Table \ref{tab:ExtModelCompare} and \ref{tab:ExtConModelCompare} in Appendix \ref{sec.APD.ExtendedModel}. 
\end{table*}

To understand the underlying causes of NeuSight’s performance patterns, we analyze its training methodology. NeuSight uses the final latency value serving as the target in the loss function. This approach introduces two key issues:
\begin{itemize}
    \item \textbf{Imbalanced Loss Sensitivity}: Latency values vary widely across devices and configurations. Loss functions such as Symmetric Mean Absolute Percentage Error (SMAPE) or Mean Absolute Relative Error (MARE) are sensitive to this imbalance. Small errors on low-latency samples can disproportionately inflate the loss, causing the model to prioritize accuracy on fast kernels while neglecting slower ones.
    \item \textbf{Device-Specific Bias}: More powerful devices like the A100 yield significantly lower latency for the same input/output dimensions. As a result, NeuSight tends to prior to low-latency samples, reducing its generalization across devices. For example, reducing from 70 to 35 µs may significantly lower the loss, even though the impact on overall latency is minimal.
\end{itemize}

However, when switching to BF16 datatype, NeuSight limitation on differentiating the performance disparities among a vast number of algorithm combinations becomes a dominant factor behind its enormous worst-case error rate, as discussed in subsection \ref{subsec.Eval.PerLayer}. Notably, we observe a reverse trend compared to FP32: NeuSight’s error rates on A100 and 5070 become the highest among all five devices. This behavior is partly due to the smaller ground-truth latency values in BF16, which make relative-error-based loss functions such as SMAPE and MARE more sensitive to unseen factors. In contrast, our proposed method continues to deliver stable predictions, with errors mostly below 10\%. 

We can also identify an abnormal behavior on the NVIDIA A100 device: with the Qwen-3 0.6B model, a batch size of one requires 47ms to complete, whereas a batch of eight takes 91ms. Interestingly, when running with batch sizes of two or four, A100 still needs about 47ms per batch. This anomaly suggests that the GPU is not fully utilizing its parallelism for smaller batch sizes, resulting in inefficient scaling. The lack of sufficient parallel work and its superior computational power on BF16 prevents the hardware from saturating its compute resources, and the fixed kernel-launch overhead becomes a dominant factor in execution time \cite{nvidia2023gpuperf}. As a result, latency does not decrease proportionally with smaller batches on small models, and the expected throughput gains are not realized, resulting an error rate of 51.4\% with PM2Lat. These observations suggest that with smaller datatype, we need to carefully analyze the parallelism saturation problem. 

\begin{table}[t]
    \centering
    \caption{Model-Wise Comparison (cont). DS R1: DeepSeek R1 Distill Qwen. Other devices not listed as OOM happens.}
    \label{tab:ModelCompareCont}
    \begin{tabular}{|c|c|ccc|ccc|}
    \hline
        & & \multicolumn{3}{|c|}{L4} & \multicolumn{3}{|c|}{A100} \\
        Model& BS&MeanT&PL&NS&MeanT&PL&NS\\
        &&(ms)&(\%)&(\%)&(ms)&(\%)&(\%)\\ \hline
        &1&186&\textbf{\color{red}+3.1}&+47.8&49&\textbf{\color{red}+7.0}&+152.3\\
        DS R1&8&1466&\textbf{\color{red}+1.1}&+49.1&335&\textbf{\color{red}+0.6}&+183.5\\
        7B&16&2974&\textbf{\color{red}+1.3}&+54.3&645&\textbf{\color{red}+2.8}&+179.4\\
        &32&&&&1259&\textbf{\color{red}+4.4}&+175.7\\ \hline 

        %%%%%%%%%%%%%%%%%%%%%%%%%%%%%%%%%%%%%%%%%%%%%%%%%%%%%%%%%%%%%%%%%%%%%%
        
        &1&-&-&-&92&\textbf{\color{red}+13.1}&+152.4\\
        DS R1&8&-&-&-&669&\textbf{\color{red}+0.1}&+157.8\\
        14B&16&-&-&-&1314&\textbf{\color{red}+0.2}&+161.9\\ \hline
    \end{tabular}
    %An extended version with more models can be found in Table \ref{tab:ExtModelCompare} and \ref{tab:ExtConModelCompare} in Appendix \ref{sec.APD.ExtendedModel}. 
\end{table}

The results from model-level latency prediction confirm that PM2Lat is not only accurate at the layer level but also highly effective when applied to full DNN models. Its consistent performance across diverse architectures, data types and hardware demonstrates strong generalizability. Unlike NeuSight, which suffers from device-specific biases and loss function sensitivity, PM2Lat maintains low error rates without requiring extensive training or tuning. Its interpolation-based approach and kernel-specific profiling allow it to adapt to a wide range of input/output dimensions, avoiding overfitting problem. 

However, PM2Lat can predict latency for only static DNN topologies, which have fixed input/output features and varying batch sizes—similar to NeuSight and other SoTA methods. When dealing with architectures that involve dynamic computation flows, such as Mixture-of-Experts (MoE), where each token activates only a subset of experts, a more detailed analysis is required. Specifically, it is necessary to accurately predict how many experts will be activated and how many tokens will be routed to each expert. Once this information is available, PM2Lat can still be applied to estimate execution time. Addressing this challenge is left for future work.

%Moreover, PM2Lat’s prediction speed—over 140× faster than NeuSight—makes it particularly well-suited for large-scale NAS and deployment scenarios involving multiple devices. The ability to reduce preprocessing time from hundreds of days to just a few highlights its practical value. These advantages position PM2Lat as a scalable, efficient, and reliable solution for latency-aware model design and deployment in real-world systems.

\subsection{PM2Lat on other Custom kernels} \label{subsec.Eval.Others}

\begin{table}[]
    \centering
    \caption{Error of PM2Lat on custom kernels.}
    \label{tab:ExtendPM2Lat}
    \begin{tabular}{|c|c|c|c|c|c|c|}
        \hline
        &&3060M&T4&L4&A100&5070\\\hline
        TritonMM & PL & 6.9 & 4.4 & 5.7 &9.6& 4.8\\
        & PL TruthCFG & 6.4 & 5.2 & 7.1 & 8.2&4.3\\
        \hline
        TritonVec & PL&0.8&1.7&5.6&4.0&7.4\\\hline
        F-Attn & PL&3.1&-&7.4&5.6& -\\\hline
        C-Attn & PL&5.7&7.9&5.7&4.0&-\\\hline
    \end{tabular}\\
    PL TruthCFG: running PM2Lat with the target optimal configuration resulted by Triton).\\
    F-Attn: Flash Attention. C-Attn: Cutlass Attention.
\end{table}
As discussed in previous sections, MatMul serves as a representative kernel for analyzing the limitations of linear regression in latency prediction. The insights gained from this analysis guide the development of PM2Lat’s interpolation-based strategy, which we now extend to other computation-intensive kernels. These include custom kernels written in Triton and fused attention mechanisms, e.g. Triton MatMul, Flash Attention and Cutlass Attention. Although convolutional layer are also computation-intensive, its complexity is beyond MatMul layers (e.g. with Fast Fourier Transform, Winograd algorithm). Hence we, will leave it for future work.

%A detailed analysis of these custom GPU kernels is provided in Appendix~\ref{sec.APD.CustomGPU.CutlassAttention}. 
Overall, PM2Lat applies the interpolation strategy with kernel-level differentiation originally developed for MatMul (Section~\ref{sec.ProposedMethod}) as a foundation, adapting it with kernel-specific data collection resolutions and profiling strategies. Table \ref{tab:ExtendPM2Lat} demonstrates PM2Lat's error rate on custom GPU kernels. Across five GPU models---RTX 3060M, T4, L4, A100 and RTX5070---PM2Lat consistently achieved low prediction errors. For TritonMM, errors ranged from 4.4\% to 9.6\%, with slightly improved accuracy when using Triton’s optimal configuration (PL TruthCFG). 
%TritonVec---vector-based layer implemented by Triton--- showed particularly strong performance, with errors as low as 0.8\% on 3060M and under 8\% on all GPUs. 
Attention kernels also demonstrated solid results, with Flash Attention and Cutlass Attention yielding errors between 3.1\% and 7.9\%, depending on the GPU. Please note that we are using FlashAttention v2, which does not support Tesla Turing–based GPUs such as the NVIDIA T4. In addition, neither attention method currently supports the Blackwell architecture (RTX 50xx series). In conclusion, these findings highlight PM2Lat’s robustness and adaptability to diverse kernel types and hardware platforms.

\subsection{Applications of PM2Lat}\label{subsec.Eval.Applications}
This subsection introduces two representative applications that demonstrate the practical impact of PM2Lat.
%First, we employ PM2Lat as a foundational latency prediction method for optimization problems in distributed DNN models, enabling efficient workload allocation across heterogeneous environments. Second, we integrate PM2Lat into NAS preprocessing for resource-constrained devices, where accurate latency estimation is essential for balancing performance and hardware limitations. These applications not only validate the versatility of PM2Lat but also highlight its role in improving system efficiency and scalability in real-world scenarios.

\subsubsection{Resource Management/Allocation/Optimization in Distributed DNN Models}\label{subsubsec.Eval.Applications.DDNN}
Resource management and allocation in distributed DNN models is a critical challenge due to the heterogeneity of devices and the complexity of workload distribution. %Efficient optimization requires accurate latency prediction to guide decisions such as layer placement, parallelization, and communication scheduling. 
In this context, we distribute the Qwen3-4B model across two GPUs: 3060M and 5070, with the input arriving at 3060M first. The model is executed with a batch size of eight, which exceeds the memory capacity of either devices when running the entire model independently. As noted in \cite{le2024optimal} and \cite{zhang2024edgeshard}, distributing a model across devices and applying pipeline parallelism helps reduce idle time.
However, the overall system performance is constrained by the stage with the highest execution time. Minimizing this bottleneck is therefore essential for improving throughput and efficiency. While prior work employs dynamic programming for partitioning across multiple devices, our scenario involves only two devices, resulting in a single partition point. Consequently, the optimal strategy is to heuristically select the partition that minimizes the maximum execution time across both devices.

With PM2Lat as the core prediction method, the system recommends a partition point immediately after the 12th Transformer block, whereas with NeuSight, it selects the point after the 13th block. Furthermore, with PM2Lat, it predicts a bottleneck latency of 570$ms$, compared to 951$ms$ with NeuSight. After applying these partitioning plans to process 100 requests, the system with PM2Lat completes these requests in 58.1$s$, while with NeuSight it requires 62.2$s$. Although a difference of 4.1$s$ may appear small, even minor delays can accumulate into significant performance degradation when the system operates continuously over extended periods, as noted in Section \ref{sec.Intro}. Moreover, NeuSight’s bottleneck estimation deviates by approximately 52.9\% (951$ms$ vs. 622$ms$ per request), which is a substantial error. Such inaccurate predictions can lead applications that aim to optimize the trade-off between latency and accuracy toward suboptimal architectures. In other words, in most cases, the architecture would need to be aggressively reduced to meet latency constraints, potentially sacrificing accuracy unnecessarily. 

Furthermore, in applications that prioritize energy consumption or aim to optimize inference cost, low-accurate latency estimation can lead to inefficient resource allocation, higher operational expenses, or inappropriate budget planning \cite{shekhar2024towards}. 
%This is because both energy consumption and inference cost are largely determined by execution time. %PM2Lat’s accurate predictions help mitigate these risks by enabling balanced decisions that account for performance, cost, and resource constraints.
%{CP version here}
In detail, we already know that the energy consumption of a kernel during execution is governed by the basic physical relationship:
\begin{equation}
    E_{\text{kernel}}=P_{\text{kernel}}\times t_{\text{kernel}}
\end{equation}
where $E_{\text{kernel}}$ is the total energy used by the kernel, $P_{\text{kernel}}$ is the average power draw of the underlying compute units while the kernel is running, and is its execution time. Because power tends to remain relatively stable for a given hardware state---especially within a single kernel’s execution phase under SIMT architecture---the runtime $t_{\text{kernel}}$ becomes the primary driver of energy consumption. Consequently, inaccuracies in latency estimation at the kernel level can propagate directly into proportionally inaccurate energy calculations. However, power measurement or prediction is not within the scope of this paper. Consequently, we do not evaluate the application of PM2Lat in minimizing energy consumption, and we leave the integration of power models—together with the results presented in the following section—to a NAS‑based framework that constructs models tailored to the target device by jointly minimizing latency and energy consumption while accounting for accuracy trade‑offs as future work.
%This tight coupling underscores the importance of precise per-kernel latency modeling: errors in predicting $t_{\text{kernel}}$ translate linearly into errors in $E_{\text{kernel}}$ affecting both model-level energy estimates and higher-level optimization decisions.

\subsubsection{NAS Preprocessing for resource-constrained devices}\label{subsubsec.Eval.Applications.NASpre}
Nowadays, designing DNNs/LLMs for deployment on resource-constrained devices \cite{akhauri2024latency,cai2018proxylessnas} has become increasingly important due to the growing demand for real-time, on-device intelligence. 
%Efficient deployment is crucial to achieve low-latency inference while preserving model accuracy, enabling reliable performance in applications such as autonomous systems, mobile platforms, and IoT environments. Hence, it is common to address the trade-off between latency and accuracy of the target model. 
In this scenarios, it is common to address the trade-off between latency and accuracy of the target model. 
To optimize this balance, NAS is often employed to guide the design of model topologies under strict resource constraints. However, the sheer number of possible architectures explored during NAS makes direct latency measurement impractical, as evaluating each candidate on real hardware is prohibitively expensive. 
%the need for deploying transformers on Res Con .... 
For example, in a Transformer model with 14 choices for input/output features, batch sizes ranging from 1 to 256, and sequence lengths from 64 to 8192, the number of configurations for just one MatMul layer exceeds 400 million possibilities.
%Ok, we should use NAS and predict the latency of each possible config  ... to find the best ...
Therefore, latency prediction models are often employed to estimate inference time efficiently.
%, enabling rapid pruning of suboptimal designs without exhaustive benchmarking. 
%However, the processing time required for each prediction also impacts the overall duration of the NAS process, especially when evaluating a large number of candidate architectures. 
Especially, we usually pre-process all the possibilities and cache them for future reuse.
PM2Lat, with its advantages of lightweight and accuracy, serves as a promising baseline for latency prediction in NAS  workflows. By reducing computational overhead while maintaining reliable estimates, PM2Lat enables faster architecture evaluation and accelerates the overall search process without compromising prediction quality. 

To quantify this, we perform 1000 predictions using both PM2Lat and NeuSight, and record the processing time. The results show that NeuSight requires an average of 6.5$ms$ per prediction (GPU-based), whereas PM2Lat needs only 0.045$ms$ per prediction (CPU-based). Hence, with the example above, NeuSight would take approximately 30 days to process and cache predictions for MatMul layers alone, while PM2Lat completes the same task in about five hours. This big difference highlight PM2Lat’s suitability for large-scale NAS preprocessing, where efficiency and accuracy are equally critical. Once the preprocessing data is available, the problem transitions into an optimization task, as demonstrated in subsection \ref{subsubsec.Eval.Applications.DDNN}. Therefore, we do not perform a full NAS process here.

\section{Conclusion}\label{sec.Conclusion}

We proposed PM2Lat, a fast, accurate, and generalized framework for latency prediction of DNNs on NVIDIA GPUs by leveraging the SIMT architecture and analytically modeling kernel throughput. 
%, PM2Lat demonstrates that interpolation-based techniques can achieve high accuracy without the computational overhead of deep learning models. 
Our evaluation shows that PM2Lat consistently outperforms NeuSight, achieving less than 10\% error on modern Transformer models and maintaining a low error rate of 3–8\% across a diverse set of GPU kernels. Furthermore, PM2Lat exhibits strong generalizability across data types, layer types, and hardware platforms, making it a practical solution for latency-sensitive applications such as distributed inference and NAS preprocessing.
%PM2Lat’s design enables practical deployment in latency-sensitive systems, large-scale neural architecture search, and multi-device inference environments. 
%Furthermore, by retaining the strategic foundation of latency-aware scheduling used in distributed training frameworks like NeuSight, PM2Lat offers a scalable alternative that replaces the core prediction engine with a more efficient analytical model. Its extensibility to custom kernels and emerging frameworks positions PM2Lat as a robust tool for modern AI workloads. 
Future work will extend PM2Lat to support convolutional and recurrent layers, as well as custom operators used in different frameworks like llama.cpp, aiming to provide comprehensive latency modeling across the full spectrum of DNN architectures.

\section*{Acknowledgment}
This work was supported in part by the Norwegian Research Council under Grant 322473 (AirQMan project) and the European Commission under Grant 101086541 (MISO project).

\printbibliography

\end{document}